\documentclass[12pt]{article}
\usepackage{graphicx}
\usepackage{amsmath}

\usepackage{amscd,amssymb}
\usepackage[cmtip,arrow,
matrix]{xy}

\usepackage{curves}

\newcommand{\cE}{\mathcal{E}}
\newcommand{\cF}{\mathcal{F}}

\newcommand{\cG}{\mathcal{G}}

\newcommand{\id}{{1\!\!1}}

\DeclareMathOperator{\Hom}{Hom}

\def\cR{{\mathcal R}}
\def\cS{{\mathcal S}}

\def\cA{{\mathcal A}}

\def\cC{{\mathcal C}}

\def\cD{{\mathcal D}}

\def\cH{{\mathcal H}}
\def\cP{{\mathcal P}}
\def\cT{{\mathcal T}}

\def\cE{{\mathcal E}}
\def\cF{{\mathcal F}}
\def\cX{{\mathcal X}}

\def\cG{{\mathcal G}}
\def\cO{{\mathcal O}}
\def\cQ{{\mathcal Q}}

\def\mL{{\mathfrak L}}
\def\mg{{\mathfrak g}}
\def\mh{{\mathfrak h}}
\def\ml{{\mathfrak l}}
\def\Inv{{\rm Inv}}
\def\Ext{{\rm Ext}}

\newtheorem{theorem}{Theorem}[section]
\newtheorem{lemma}{Lemma}[section]

\newtheorem{conjecture}{Conjecture}[section]

\newtheorem{definition}{Definition}[section]

\newtheorem{remark}{Remark}[section]

\begin{document}


\vspace{.1in}

\begin{center}

{\Large\bf BRST-Invariant Deformations of Geometric Structures in Topological Field Theories}

\end{center}
\vspace{0.1in}

\begin{center}
{
A. A. Bytsenko $^{(a)}$
\footnote{aabyts@gmail.com}},
M. Chaichian $^{(b)}$
\footnote{masud.chaichian@helsinki.fi}, 
A. Tureanu $^{(b)}$
\footnote{anca.tureanu@helsinki.fi}
and F. L. Williams $^{(c)}$
\footnote{williams@math.umass.edu}

\vspace{7mm}
$^{(a)}$ 
{\it
Departamento de F\'{\i}sica, Universidade Estadual de
Londrina, Caixa Postal 6001,
\\ Londrina-Paran\'a, Brazil}

\vspace{0.1mm}
$^{(b)}$
{\it
Department of Physics, University of Helsinki,
P.O. Box 64, FI-00014 Helsinki, Finland}

\vspace{0.1mm}
$^{(c)}$
{\it
Department of Mathematics and Statistics, University of Massachusetts \\
Lederle Graduate Research Tower 710 North Pleasant Street
Amherst, MA 01003, USA}

\vspace{3mm}

\end{center}

\vspace{0.1in}
\begin{center}
{\bf Abstract}
\end{center}
We study a Lie algebra of formal vector fields $W_n$ with its application to the perturbative 
deformed holomorphic symplectic structure in the A-model, and a Calabi-Yau manifold with 
boundaries in the B-model. A relevant concept in the vertex operator algebra and the BRST cohomology is that of the elliptic genera (the one-loop string partition function). We show that the elliptic genera can be written in terms of spectral functions of the hyperbolic three-geometry (which inherits the cohomology structure of BRST-like operator). We show that equivalence classes of deformations are described by a Hochschild cohomology theory of the DG-algebra ${\mathfrak A} = (A, Q)$, $Q =\overline{\partial}+\partial_{\rm deform}$, which is defined to be the cohomology of $(-1)^n Q +d_{\rm Hoch}$. Here $\overline{\partial}$ is the initial non-deformed BRST operator while $\partial_{\rm deform}$ is the deformed part whose algebra is a Lie algebra of linear vector fields ${\rm gl}_n$. We discuss the identification of the harmonic structure $(HT^\bullet(X); H\Omega_\bullet(X))$ of affine space $X$ and the group ${\rm Ext}_{X²}^n(\cO_{\triangle}, \cO_{\triangle})$ (the HKR isomorphism), and bulk-boundary deformation pairing.

\vfill



\newpage
\tableofcontents

\renewcommand{\thefootnote}{\arabic{footnote}}
\setcounter{footnote}{0}

\section{Introduction}

Two-dimensional topological field theories have been actively studied (especially in connection with mirror symmetry), however these theories are much more nontrivial and more interesting when can be defined for deformed holomorphic symplectic structure of the cotangent bundle
(the A-model) and a Calabi-Yau manifold with boundaries (the B-model).
Recall that original topological field model in arbitrary dimension has been analyzed by M. 
Atiyah \cite{Atiyah} and this theory did not allow for boundaries (but it allowed for defects 
of higher codimension, Wilson loops in Chern-Simons theory, for example).
Non-trivial variety of boundary conditions which could be associated with topological branes 
in the B-model has been introduced by E. Witten \cite{Witten}.
\footnote{
In the topological B-model boundary conditions can be twisted by the curvature of the 
Chan-Paton factors and the $B$ field \cite{Abouelsaood,Callan} 
}

In this paper we study (perturbative) deformations of complex structure for topological sigma models. There are strong reasons to study these models. A generalization of topological field theories is most conveniently formulated in the language of category theory. In two-dimensional case the structures are provided by A- and B-models associated to Calabi-Yau manifolds \cite{Witten,Douglas} (to Calabi-Yau $A_\infty$ categories \cite{Costello,KontsevichS} for more abstract cases). The topological A- and B-twist respectively capture the symplectic and holomorphic structure on the target Calabi-Yau manifold. Mirror symmetry interchanges the topological A-model on the half of a mirror pair with the B-model on the other.
Homological mirror symmetry formulated by M. Kontsevich \cite{Kontsevich95} postulates a 
quasi-equivalence between certain $A_\infty$-categories related to the A- and B-model. Indeed, to the A-model one associates the Fukaya category (some yet unknown generalizations thereof), and to the B-model one associates the bounded derived category of coherent sheaves.
A mathematical reason to study topological field theory is its path intergal analysis which suggests the Fukaya-Floer category of  boundary conditions in the A-model, and deformation quantization and the derived category of coherent sheaves on a complex manifold in the B-model. 

A brief summary of the results of the paper is the following.
We study a Lie algebra of formal vector fields $W_n$ 
\footnote{The concept of {\it formal geometry} has been introduced in \cite{Gelfand1,Gelfand2}, where by means of calculation of relative cohomologies of the Lie algebra of formal vector 
fields on manifold the characterictic classes of its tangent fibre bundles has been constructed.}
with it application to the deformed BRST operator. We consider the infinite-dimensional Lie algebra
$W_n\ltimes {\overline G}\otimes P_n$, i.e. a semidirect sum of algebra $W_n$ extended by product of Lie algebra $\overline G$ 
(related to the initial non-deformed BRST operator $\overline{\partial}$) and formal power series $P_n$ of $n$ compex variables $z_1, \ldots, z_n$.
The infinite-dimensional Lie algebra $W_n\ltimes {\overline G}\otimes P_n$ contains a small Lie subalgebra ${\rm g}{\rm l}_n\oplus {\overline G}$, which consists of linear {\it vector fields} and ${\overline G}$-valued fields. 
The cochain complex of a Lie algebra $W_n\ltimes {\overline G}\otimes P_n$ is isomorphic to the quotient of Weyl algebra of ${\rm g}{\rm l}_n\oplus {\overline G}$ by $(2n+1)$-st term of standard filtration. 
There are morphisms between filtered complexes and spectral sequences of these two Lie algebras. These subtleties have the effect of realizing spectral sequences directly in BRST cohomology of the operator 
$
Q = \overline{\partial} + \partial_{\rm deform}
$
(Lie algebra of the operator $\partial_{\rm deform}$ is ${\rm gl}_n$). $Q$ satisfies the 
condition $Q^2 = 0$ and we show that on the bigraded vector space $C^\bullet(A)$ the 
Hochschild cohomology of an associative algebra $A$ is defined to be the cohomology of 
$(-1)^n (\overline{\partial} + \partial_{\rm deform})+d_{\rm Hoch}$. We also discuss the 
identification of the harmonic structure $(HT^\bullet(X); H\Omega_\bullet(X))$ of (affine) 
space $X$ and the group ${\rm Ext}_{X²}^n(\cO_{\triangle}, \cO_{\triangle})$ 
(the Hochschild-Kostant-Rosenberg isomorphism (HKR)) and bulk-boundary 
deformation pairing.

This paper is outlined as follows. In Section 2 we analyze two-categories of topolgical 
sigma-models and deformations
of holomorphic symplectic structures in the A-model, and descent procedure of deformations of 
boundary conditions in the B-model.
In that section we mainly follow the lines of \cite{Kapustin08,Kapustin09}. 
We discuss Lie algebras of formal vector fields in Section 3 and then, in Section 4, we describe tensor modules and invariants of Lie algebra $W_n\ltimes {\overline G}\otimes P_n$, its Lie subalgrebra
${\rm g}{\rm l}_n$, and truncated Weyl algebras. Filtrations on cochain complexes and morphisms of filtered complexes are analyzed in Section 5, and then in Section 6 we discuss the BRST cohomology and characteristic classes of foliations with its connection to the Charn polynomials (and elliptic genera). We show that the final result can be written in terms of spectral functions of the hyperbolic three-geometry associated with $q$-series.
The Hochschild cochain complex and deformations are studied in Section 7. Finally deformation pairing, algebra deformations and a bulk-boundary operator product expansion (OPE) are analyzed in Section 8. In the appendixes A, B and C we deduce the useful information on Lie algebra cohomology, spectral sequences, sheaves and categories, and the HKR isomorphism.

\section{Two-categories of sigma-models and their deformations}
\label{Two-categories}
{\bf The general case.}
In this Section we will mostly follow \cite{Kapustin08,Kapustin09} in the reproduction of necessary results.
Let ${\cX} =(X,s)$ be a pair in which $X$ is a real manifold and $s$ is a geometric structure on $X$ (such as a complex structure or a symplectic structure). Let $\Sigma$ be a real $N$-dimensional manifold. A topological sigma-model with world-volume $\Sigma$ and a target space $\cX$ is a quantum field theory based on a path integral (the measure on the space of maps $\Sigma \rightarrow X$ is dertermined by the structure $s$). As $\cX$ varies, one obtains an
$N$-category with useful features, where the $X$ component of $\cX$ serves as a target space for a topological sigma model. For the definition of categories see Appendix B.

Recall that the category $\cC$, associated with pair $(X, s)$, has a symmetric monoidal structure related to the cartesian product of manifolds:
$
\cC\times\cC\rightarrow\cC,\,
(X_1, s_1)\times(X_2,s_2) = (X_1\times X_2, s_1\times s_2),
$
with the natural structure $s_1\times s_2$  on $X_1\times X_2$.
This monoidal structure has a unit element $\cX_{pt}=(X_{pt}, s_{pt})$, where $s_{pt}$
is the corresponding trivial structure of the manifold consisting of a single point $X_{pt}$. It is clear that
$\cX_{pt}\times \cX= \cX$. Define a $(N-1)$-category of morphisms 
$\cC := {\rm Hom}_{\cC}(\cX_{pt}, \cX)$.
In quantum field theory this categogy is known as the category of boundary conditions of the topological sigma-model associated with $\cX$. The category $\cC$ has a contravariant
duality functor
$
\cC \stackrel{\lozenge}{\rightarrow} \cC,\,
(X,s)^{\lozenge} = (X, s^{\lozenge}),
$
such that there is a canonical equivalence between $(N-1)$-categories
of morphisms:
$
{\rm Hom}_{\cC}(\cX_1,\cX_2) = {\rm Hom}_ {\cC} (\cX_{pt}, \cX^{\lozenge}_1\times \cX_2)
:=\cC_{\cX_1^{\lozenge}\times \cX_2}.
$
This equivalence implies that an object
${\mathfrak E}_{12}\in {\rm Hom}_{\cC}(\cX_1,\cX_2)$ determines a functor between $(N-1)$-categories
$
F [{\mathfrak E}_{12}]: \,\cC_{\cX_1} \rightarrow \cC_{\cX_2},
$
which represents a composition of morphisms within $\cC$.
Moreover, a composition of morphisms of $\cC$ corresponds to the composition of functors, so the structure of the $N$-category $\cC$ is
determined by the boundary condition categories $\cC_{\cX}$ and the
functors.
\\
\\
\noindent
{\bf The differential Gerstenhaber algebra and deformations of the complex structure.}
For the further discussion we need the differential Gerstenhaber algebra $\cG(X)$ which gives rise to a Kuranishi deformation theory \cite{Clemens,Joyce} for any complex structure. Let $\cR$ be a ring with unit and let $\cA$ be an $\cR$-algebra. Suppose that ${\mathfrak a} =\oplus_{n\in {\mathbb Z}}{\mathfrak a}^n$ is a graded algebra over $\cA$. If $a\in {\mathfrak a}^n$, let $|a|$ denote its degree. 
The algebra ${\mathfrak a}$ is called a Gerstenhaber algebra \cite{Gerstenhaber,Mackenzie} if there is an associative product $\wedge$ and a graded commutative product $[-\bullet -]$ 
satisfying the following axioms: for $a\in {\mathfrak a}^{|a|}$, $b\in {\mathfrak a}^{|b|}$,  $c\in {\mathfrak a}^{|c|}$\,,
\begin{eqnarray}
\!\!\!\!\!\!\!\!\!
&& a\wedge b \in {\mathfrak a}^{|a|+ |b|},\,\,\,\,\,\,\, b\wedge a = 
(-1)^{|a||b|}a\wedge b\,,
\nonumber \\
\!\!\!\!\!\!\!\!\!
&& (-1)^{(|a|+1)(|c|+1)}[[a\bullet b]\bullet c ] \, + \, 
{\rm more}\,\,\,{\rm two}\,\,\,\,{\rm terms}\,\,\,\,(a, b, c \,\,\,{\rm cyclic}\,\,\,{\rm permutation}) = 0\,, 
\nonumber \\
\!\!\!\!\!\!\!\!\!
&& [a\bullet b\wedge c] = [a\bullet b]\wedge c + (-1)^{(|a|=1)|b|}
b\wedge [a\bullet c]\,.
\end{eqnarray}
Note that a differential graded algebra ${\mathfrak a}$ is a graded algebra with a graded commutative product $\wedge$ and a differential $d$ of degree +1, i.e. a map $d: {\mathfrak a}\rightarrow  {\mathfrak a}$ such that
\begin{equation}
d({\mathfrak a}^n)\subseteq {\mathfrak a}^{n+1},\,\,\,\, 
d\circ d = 0,\,\,\,\, 
d(a\wedge b)= da\wedge b + (-1)^{|a|}a\wedge db\,.
\end{equation}
Let ${\mathfrak a}$ be a graded algebra over $\mathbf k$ such that $({\mathfrak a}, [-\bullet -], \wedge)$
forms a Gerstenhaber algebra and $({\mathfrak a}, \wedge, d)$ forms a differential graded algebra. If in addition
\begin{equation}
d[a\bullet b] = [ad\bullet b] + (-1)^{|a|+1}[a\bullet db]\,,
\,\,\,\,\,\forall\, a,b \in {\mathfrak a}\,,
\end{equation}
then $({\mathfrak a}, [-\bullet -], \wedge, d)$ is a differential Gerstenhaber algebra.

Let $X$ be a compact K\"ahler manifold with the K\"ahler form 
$\omega$; in physics, the mathematical setting is restricted to D-branes on Calabi-Yau manifolds. Let us consider the decomposition
$
({\mathfrak g}\oplus {\mathfrak g}^*)\otimes {\mathbb C}:=(TX\oplus T^\vee X)\otimes {\mathbb C} \equiv E\oplus E^\vee$, where 
$T^\vee X$ is a cotangent bundle,  
${\mathfrak g}$ and ${\mathfrak g}^*$ are complex Lie subalgebras of the complexified algebra ${\mathfrak g}^{\mathbb C}$.
The differential Gerstenhaber algebra 
$(\Omega^\bullet E^\vee, \overline{\partial}_{E^\vee})$ is elliptic and it gives rise to a Kuranishi deformation theory for any generalized complex structure \cite{Goto}. The space of infinitesimal deformations of complex structures on $X$ is given by the direct sum of the ${\mL}^{-1}_{s}$-valued Dolbeault cohomology groups \cite{Goto,Bytsenko}:
\begin{equation} \label{deformations}
\cH^{n,2}_{\overline{\partial}}(X; \mL^{-1}_{s})\oplus \cH^{n-1,1}_{\overline{\partial}}(X; \mL^{-1}_{s})\oplus \cH^{n-2,0}_{\overline{\partial}}(X; \mL^{-1}_{s}),
\end{equation}
where $\mL^{-1}_s$ denotes the dual of the standard canonical line bundle of the complex manifold $(X, s)$, and where we emphasize 
$\overline{\partial}$ cohomology by the subscript $\overline{\partial}$. The spaces of infinitesimal deformations  of complex structures are:

The space  $\cH^{n,2}_{\overline{\partial}}(X; \mL^{-1}_{s})$ given by the action of {B}-fields {\rm (}$2$-forms{\rm )}\,.

The space $\cH^{n-2,0}_{\overline{\partial}}(X; \mL^{-1}_{s})$  induced by the action of holomorphic $2$-vector fields\,.

The space of the obstructions given by
\begin{equation} \label{obstructions}
\cH^{n,3}_{\overline{\partial}}(X; \mL^{-1}_{s})\oplus \cH^{n-1,2}_{\overline{\partial}}(X; \mL^{-1}_{s})\oplus \cH^{n-2,1}_{\overline{\partial}}(X; \mL^{-1}_{s})\oplus \cH^{n-3,0}_{\overline{\partial}}(X; \mL^{-1}_{s})\,.
\end{equation}
As it follows from Eqs. (\ref{deformations}) and (\ref{obstructions}) the space $\cH^{n-1,1}_{\overline{\partial}}(X; \mL^{-1}_{s})\cong H^1(X, \cO)$ is the space of infinitesimal deformations of complex structures in Kodaira-Spencer theory. 

\noindent
\begin{remark}
Recall that the general Kodaira-Spencer theory identifies as the key to the deformation theory the sheaf cohomolgy group $H^1(X, \cO)$, where $\cO$ is the sheaf of germs of sections of the holomorphic tangent bundle. Some results on infinitisimal deformations of complex structure are: 

{\bf (i)} \, The base of the Kuranishi family lies in $H^2(X)= \oplus_{p+q=2}H^q(X, \Omega^pT^\vee X)$. 

{\bf (ii)} \,The image of the obstruction map lies in $H^3(X)= \oplus_{p+q=3}H^q(X, \Omega^pT^\vee X)$. 

{\bf (iii)} In the deformation theory any deformation has three components $(\xi_0, \xi_1, \xi_2)$, namely
\begin{equation}
\xi_0 \in H^0(X, \Omega^2T^\vee X),\,\,\,\,\,\,\,
\xi_1 \in H^1(X, T^\vee X),\,\,\,\,\,\,\, 
\xi_2 \in H^2(X, \cO)\,.
\label{deform11}
\end{equation}
The component $\xi_0\in C^{\infty}(\Omega^2T^\vee X)$ is a type of deformation for complex manifolds for which the integrability condition is simply that
$$
\overline{\partial}\xi_0 + (1/2)[\xi_0, \xi_0]_{\overline{\partial}}=0
\,\,\,\,\,\,\,{\rm(Maurer-Cartan}\,\,\, {\rm equation).}
$$ 
This condition is satisfied if and only if the bivector $\xi_0$ is holomorphic and Poisson. 
The component $\xi_1$ is a deformation of the complex structure. 
The component $\xi_2$ is a complex {\rm B}-field action.
\end{remark}

\subsection{The A-model and deformations of holomorphic symplectic structures}

Let $\cC$ be a $(N=2)$-category. For the A-model the structure $s$ is a symplectic structure (that is, $s$ is a symplectic form on $X$), the category of boundary conditions $\cC_{\cX}$ is the Fukaya-Floer category $F(\cX)$, its simplest objects being lagrangian submanifolds of $X$, the action of the duality functor is $s^{\lozenge}=- s$, and the functor $F [{\mathfrak E}_{12}]$ is the lagrangian correspondence functor determined by a lagrangian submanifold $M_{12}\subset\cX_1^{\lozenge}\times \cX_2$.

Deformations of holomorphic symplectic structure which preserve the de Rham cohomology class of $\omega$ 
\footnote{
Recall that if $X$ is a symplectic manifold then it is equipped with a 2-form $\omega$ which is closed ($d\omega = 0$) and satisfies a certain non-degeneracy condition: for each $z\in X$ the bilinear form on the tangent space $T_zX$ determined by $\omega$ is non-degenerate. If $f\in C^\infty(X)$ is a smooth function, the Hamiltonian vector field generated by $f$ is the unique vector field $H_f$ such that $df = \iota (H_f)\omega$, where $\iota (H_f)$ is the contraction by the Hamiltonian vector field.
}
are parameterized up to gauge equivalence by Maurer-Cartan elements of the differential Poisson algebra $\cP (X, \omega)$.
The differential Poisson algebra $\cP (X, \omega)$ of a holomorphic symplectic manifold $(X, \omega)$ is defined as the algebra $\Omega^{0, q}(X)$ of $(0, q)$-forms on $X$ with the differential $\overline {\partial}$ and with Poisson bracket $\{ , \}$ coming from $\omega$. 
Let $X= T^\vee Y$, where $Y$ is a complex manifold, then we may consider a simpler version of this algebra, which we denote as $\cP_{T^\vee Y}$:
it is the algebra of $S^{i}TY$-valued $(0, q)$ forms
$\Omega^{0, q}(Y, S^{i}TY)$, where $S^i$ is the symmetric algebra (and its differential is $\overline{\partial}$). There is a natural injection
\begin{equation}
\Omega^{0, q}(Y, S^{i}TY) \hookrightarrow 
\Omega^{0, q}(T^\vee Y)
\label{injection}
\end{equation}
which turns an element of $\Omega^{0, q}(Y, S^{i}TY)$ into a $(0, q)$ differential form on $T^\vee Y$ having a polynomial dependence on fiber coordinates and restricting to zero on all fibers.
The bracket of $\cP_{T^\vee}(Y)$ is a well-defined restriction of the
Poisson bracket of $\Omega^{0, q}(T^\vee Y)$, so
the injection (\ref{injection}) becomes an injection of differential Poisson algebras 
$
\cP_{T^\vee}\hookrightarrow \cP(T^\vee Y).
$

If an element
$\zeta \in \Omega^{0, 1}(X)\subset\cG(X)$ (the differential Gerstenhaber algebra) satisfies the Maurer-Cartan equation,
then the corresponding deformation of the complex structure of $X$ is described by the Beltrami differential
$
\mu = \omega^{-1}({\partial} \zeta),
$
that is, the $(0,1)$ part of the deformed Dolbeault differential is
\begin{equation}
\overline{\partial}_{\rm new} = \overline{\partial} + \omega^{-1}(\partial \zeta)\,\llcorner\, \partial,
\label{diff}
\end{equation}
where $\llcorner$ denotes contraction on the holomorphic indices and exterior product on the antiholomorphic ones.
The symplectic form $\omega$ is replaced by
$
\omega \rightarrow \omega + d\zeta,
$
so that it remains of type $(2,0)$ relative to the new
complex structure. In the formula (\ref{diff})
we defined $\omega^{-1}:\, \Gamma(T^\vee X)\rightarrow \Gamma(TX)$
as the inverse of $\iota_{-}\,\omega$.
\\
\\
\noindent
{\bf Perturbative deformations.}
First let us consider the general case. Let the deformation of $(X, \omega)$ be perturbative, that is, if $\zeta$ is a formal power series
\begin{equation}
\zeta ({\varepsilon}) =
\sum_{j=1}^\infty \zeta_{j} \varepsilon^j,
\label{deform}
\end{equation}
then the Maurer-Cartan equation says that the leading coefficient $\zeta_{1}$ must be $\overline{\partial}$-closed, and its gauge equivalence class is determined by its Dolbeault cohomology class
$\widetilde{\zeta}_{1}\in {\rm Hom}^1_{\overline{\partial}}(X)$, while the relation
$
\overline{\partial}\zeta_2+(1/2)\{\zeta_1, \zeta_1\}=0
$
implies the integrability condition for $\widetilde{\zeta}_{1}$:
$
\{\widetilde{\zeta}_{1}, \widetilde{\zeta}_{1}\}=0.
$

Now let us consider deformations of the holomorphic symplectic structure of a cotangent bundle. 
As before $X=T^\vee Y$, we restrict ourselves to Maurer-Cartan elements $\zeta$ belonging to the subalgebra
$\cP_{T^\vee}(Y)\hookrightarrow \cP(T^\vee Y)$. Following \cite{Kapustin09} we consider only the deformations which do not deform the complex structure of the zero section $Y_0\subset T^\vee Y$,
and we impose the condition $\mu|_{Y_0}=0$ on the Beltrami
differential $\mu = \omega^{-1}(\partial \zeta)$.
Note that this condition is satisfied if $\zeta$ is at least quadratic as a function of holomorphic coordinates on fibers of $T^\vee Y$ :
$
\zeta = \sum_{j=2}^{\infty}\zeta_{i} ,
$
$
\zeta_{i}\in \Omega^{0, 1}(Y, S^iTY)\,.
$
The first two terms in this sum satisfy the equations
$
\overline{\partial}\zeta_2 = 0,
$
$
\overline{\partial} \zeta_3 + (1/2)
\{\zeta_2 , \zeta_2\} = 0.
$
As a result $\zeta_2$ is $\overline{\partial}$-closed, and its gauge equivalence class is determined by the Dolbeault cohomology class that it represents:
$
\widetilde{\zeta_2} \in H_{\overline{\partial}}^1(Y, S^2 TY),
$
$
\{\widetilde{\zeta_2}, \widetilde{\zeta_2}\}=0.
$
The injection (\ref{injection}) turns an element $\zeta \in \Omega^{0, q}(Y, S^i TY)$
into a $\overline{T}^\vee Y$-valued function
or, rather, {\it a formal power series}
on the total space of $T^\vee Y$. Denote this function by the same letter $\zeta$. The evaluation of $\zeta$ on a section of $T^\vee Y$ gives a map
$
\zeta :\, \Gamma(T^\vee Y)\rightarrow \Omega^{0, 1}(Y).
$
The restriction of the $(1,0)$ part of the differential $\partial \zeta$ of an element $\zeta\in\Omega^{0, q}(T^\vee Y)$ to the fibers of
$T^\vee Y$ determines a vertical holomorphic differential map
$
\partial_{\rm vrt}\zeta :\, \Gamma(T^\vee Y)\rightarrow \Omega^{0, 1}(Y, TY).
$
Let us consider a perturbative deformation (\ref{deform}). The
generating function becomes a formal power series
\begin{equation}
W(\varepsilon) = \sum_{j=0}^{\infty} W^{(j)}\,\varepsilon^j.
\label{deform1}
\end{equation}
Then the complex structure is deformed by the Beltrami differential
(see for detail \cite{Kapustin09})
$
\mu_W = - \partial_{\rm vrt}\zeta (\partial W)\,.
$

\subsection{The B-model and the descent procedure of deformations}

For the B-model $X$ is a Calabi-Yau manifold, $s$ is its complex structure, the category of boundary conditions $\cC_{\cX}$ is the bounded derived category of coherent sheaves
${\rm D}^b_{\rm coh}(\cX)$ (see Appendix B), its simplest objects being complexes of holomorphic
vector bundles on $X$, the duality functor acts trivially:
$s^\lozenge =s$, and the functor $F[{\mathfrak E}_{12}]$ is the
Fourier-Mukai transform corresponding to the object ${\mathfrak E}_{12}$.
Note that all suitable enough functors between the derived category of quasicoherent sheaves on two varieties, $\cX_1$ and $\cX_2$, are given by Fourier-Mukai transforms. 
Let $\mathfrak E$ is an object in ${D}^b_{\rm coh}(\cX_1\times \cX_2)$.
Given any object $\cD\in {D}(\cX_1)$ define the integral transform with kernel $\mathfrak E$ to be the functor (or morphism)
\footnote{
The following functors can be applied \cite{Aspinwall}:
{\bf (i)}\, ${\mathbf L}\pi^*_{\cX}$ - take the complex to be a complex of locally-free sheaves or, equivalently, vector bundles. $\pi^*_{\cX}$ is then the usual pull-back map on vector bundles.
{\bf (ii)}\, $\stackrel{\bf L}{\otimes}\mathfrak E$ - the complex must be a complex of locally-free sheaves; $\otimes$ is the usual tensor product of sheaves. Acting on complexes, $\otimes$ produces a double complex which can be collapsed back to a single complex in the usual way. {\bf (iii)}\, ${\mathbf R}\pi_{\cX, *}$ - the complex must be a complex of injective objects. $\pi_{\cX, *}$ is then the push-forward map for sheaves.
}
\begin{eqnarray}
&& F_{\cX_1\rightarrow \cX_2}^{\mathfrak E}: \,\,\,\,\,\,\,\,\,\,\,\,\,\,
{D}^b_{\rm coh}(\cX_1)
\longrightarrow {D}^b_{\rm coh}(\cX_2)\,,
\nonumber \\
&& F_{\cX_1\rightarrow \cX_2}^{\mathfrak E}(\cD) \, = \,
{\mathbf R}\pi_{\cX_2 *}({\mathbf L}\pi^*_{\cX_1}\cD
\stackrel{\bf L}{\otimes}\mathfrak E)\,.
\end{eqnarray}
It gives action of monodromy on the derived category \cite{Aspinwall}. The projection maps from $\cX_1\times \cX_2$ to its first and second factors is:
\begin{equation}
\xymatrix@C=6mm{
&\triangle\subset
\cX_1\times\cX_2
\,\,\,\,\,\,\,\,\,\,\,\,\,\,
\ar[dl]_{\pi_{\cX_1}}\ar[dr]^{\pi_{\cX_2}}&\\
\cX_1&&
\cX_2\\
}
\label{D00}
\end{equation}
where $\pi_{\cX_1}$ and $\pi_{\cX_2}$ are the projections from $\cX_1 \times \cX_2$ to $\cX_1$ and $\cX_2$ respectively.

{\bf The descent procedure.}
A systematic way to construct deformations of boundary conditions is the descent procedure \cite{Witten}. If $\cO$ is an even (bosonic) local observable on the boundary then its descendants ${\cO}^{(1)}\in \Omega^1(X)$ and ${\cO}^{(2)}\in \Omega^2(X)$ are defined by 
\begin{equation}
d \cO = \delta_Q \cO^{(1)} + \ldots , \,\,\,\,\,\,\,\,\,\,\,
d \cO^{(1)} = \delta_Q \cO^{(2)} + \ldots 
\end{equation}
Then in forming the topological family the generalized action becomes
$
S_{\rm bulk} \Longrightarrow S_{\rm bulk} + \gamma \int_{\Sigma}{\cO}^{(2)},
$
where $\gamma$ is a formal parameter. Because of the definition of ${\cO}^{(2)}$ it follows that the modified action is BRST-invariant up to terms proportional to equations of motion. It is clear that local topological observables on the boundary are of the same $(0, q)$-form as in the bulk. 
An element of space of infinitisemal deformations can be represented by a $\overline{\partial}$-closed inhomogeneous form $W$ of even degree. Thus the corresponding observable can be thought of as an even function $W(\phi, \eta)$ of bosonic variables $\phi^i,\,  \phi^{\overline{i}}$ and fermionic variables $\eta^i$, it satisfying 
\begin{equation}
Q\,W = \sum_i{\eta}^{\overline i}
\frac{\partial W(\phi, \eta)}{\partial {\phi}^{\overline i}} 
= 0\,.
\label{QW}
\end{equation}
Next the descendans of $W$ have to be constructed. Note that {\it on the quantum level} one requires $X$ to be a Calabi-Yau manifold, i.e. one requires the existence of a holomorphic volume form on $X$.
The volume form is used to write down a BRST-invariant measure on the space of bosonic and fermionic zero modes (see, for example, \cite{Kapustin08}).

Let $W$ be a holomorphic function on a manifold $X$. One can deform the B-model with target $X$ by modifying the transformation law for fermionic fields and adding to the action a new term which is related to this deformation (see, for detail, \cite{Kapustin08}, section Appendix B). This model can be called the Landau-Ginzburg model with target $X$ and superpotential $W$. The algebra of topological observables for the Landau-Ginzburg model is the hypercohomology of the complex
$$
\Lambda^n TX \longrightarrow \Lambda^{n-1} TX \longrightarrow \ldots \longrightarrow TX \longrightarrow \cO_X
$$
where the differential is contraction with the holomorphic 1-form
$- \partial W$. Let $X$ be a contractible open subset of
${\mathbb C}^n$ and assume that the critical points of $W$ are isolated, then the
hypercohomology of this complex is the Jacobi algebra
$
H^0(X,\cO_X)/A_{\partial W}
$
where $A_{\partial W}$ is the ideal generated by partial derivatives
of $W$. This construction can be generalized by replacing $W$ with an
inhomogeneous even $\overline{\partial}$-closed form on $X$. This generalization is called {\it the curved} B-{\it model}. 
\footnote{
To the best of our knowledge the article \cite{Guffin} was the first to study Landau-Ginzburg models over nontrivial spaces.
}
It can be thought of as a generalization of the Landau-Ginzburg deformation of the B-model \cite{Kapustin08}. 

Consider a local observable
$W(\phi,\eta)$ representing an even element of
$
\oplus_p H^{p}(\cO_X).
$
Then one can determine its descendants $W^{(j)}$ and their BRST transformations. More generally, for a non-vanishing form which is a trivial class in $\overline{\partial}$-cohomology, one can restore BRST-invariance at some order by writing 
$W{(\varepsilon)} = \sum_jW^{(j)}\varepsilon^j$ (where typically $\varepsilon^j$ is of order $\hbar^j$). Finally the total action has to be BRST-invariant. As a result we can get the curved B-model which is ${\mathbb Z}_2$-graded like the Landau-Ginzburg model.
Also the algebra of observables in the curved B-model is computed in
essentially the same way as in the Landau-Ginzburg model. Indeed,
the algebra of observables is the cohomology of a certain
differential $\delta_Q$ in the space of $(0,q)$ forms with values in
polyvector fields of type $(p,0)$. This differential is given by
\begin{equation}
\delta_Q = \overline{\partial}- \partial W\,\llcorner .
\label{deform2}
\end{equation}
If $X$ is compact and K\"ahler, then one can always find a form in the cohomology class of $W$ which is $\partial$-closed. For such $X$ the differential $\delta_Q$ reduces to $\overline{\partial}$, and the algebra of topological observables is the same as in the ordinary B-model.

\section{Lie algebras of formal vector fields}

Suppose that the even function $W(\phi, \eta)$ has the form of formal power series in fields. This situation has been discussed in Section \ref{Two-categories} for the two dimensional sigma-model (perturbative deformations of the holomorphic symplectic structure for the A-model, Eqs. (\ref{deform}), (\ref{deform1}), and the descent procedure of deformations of boundary conditions for the B-model, Eq. (\ref{deform2})). A systematic way to study deformations is to consider the space of vector fields associated with the deformed BRST operator and its topological algebra.
\\
\\
\noindent
{\bf Lie algebras $W_n^{\overline G}\ltimes {\overline G}\otimes P_n$.}
At more basic level let us consider the infinite-dimensional Lie algebra $W_n^{\overline G}\ltimes {\overline G}\otimes P_n$, i.e.
a semidirect sum of the algebra $W_n^{\overline G}$ extended by the product of the algebra $\overline G$ (related to the operator $\overline{\partial}$) and the formal power series $P_n$ of $n$ variables. These algebras are interpreted as follows:
\begin{itemize} 
\item{} The Lie algebra $W_n^{\overline G}$ is associated with the deformed part of the BRST operator and it is slightly different from the traditional infinite-dimensional Lie algebra of formal vector fields $W_n$. Its elements in ``coordinates'' can be written in the form $\sum_{i=1}^nF_i \partial/\partial z_i$, where $F_i$ is a power series in $z_1, \ldots, z_n$ whose coefficients are elements of the proper algebra $\overline{G}$ of alternating holomorphic forms. Thus when taking the commutator for such ``vector fields'' this fact should be kept in mind.
\item{} The elements of the algebra $\overline G$ correspond to the initial BRST operator \\ $\overline{\partial} =  \sum_{\overline{i}=1}^n d\overline{z}^{\overline{i}}\partial/\partial{\overline{z}^{\overline{i}}}$.
\end{itemize}
A commutator of two elements has the form
\begin{equation}
[v+g_1\otimes p_1, u+g_2 \otimes p_2]  \stackrel{{\rm def}}{=}
[v, u]_{W_n^{\overline G}} + [g_1, g_2]_{\overline G}\otimes p_1p_2 + 
g_2\otimes v(p_2) - g_1\otimes u(p_1)\,,
\label{commutator}
\end{equation}
where $v, u \in W_n^{\overline G}$, $g_i\in {\overline G}$, $p_i\in P_n$
and $i\in\{1, 2\}\,$. In the following we will concentrate on the cohomology of Lie algebras $W_n$. The corresponding theory is similar to the cohomology of Lie algebras $W_n^{\overline G}$, and the reader will have no difficulty in recovering results for the case of $W_n^{\overline G}$. The Lie algebra 
$W_n\ltimes {\overline G}\otimes P_n$ contains a Lie subalgebra
which is isomorphic to the direct sum ${\rm g}{\rm l}_n\oplus {\overline G}$. This subalgebra consists of linear vector fields and fixed ${\overline G}$-valued constant fields
\begin{equation}
\iota : \, {\rm g}{\rm l}_n\oplus {\overline G}\hookrightarrow 
W_n\ltimes {\overline G}\otimes P_n \,,\,\,\,\,\,\,\,\,\,\,
\iota (|| a_{i, j}|| + {\overline G})  =  \sum_{i, j}a_{i, j}z_i
\frac{\partial}{\partial z_j} + \sum_{\overline{i}=1}^n d\overline{z}^{\overline{i}}\frac{\partial}
{\partial{\overline{z}^{\overline{i}}}}\otimes {\id}\,.
\label{LAlgebra}
\end{equation}
\begin{remark}
Recall that if $A$ is a general field of zero characteristic then
a vector field can be writen in coordinate form $\alpha =\sum_{i=1}^nv_i\partial/\partial z_i$, where 
$v_i\in P_n = A[[z_1, \ldots, z_n]]$ is a ring of formal power series on $A^n$.
Thus an action on functions is determed by means of differentiation of formal series 
$\alpha(p)\stackrel{{\rm def}}{=}\sum_{i=1}^nv_i\partial p/\partial z_i$. The space of formal vector fields in ${A}^n$ with the projective limit topology is a topological Lie algebra with respect to the commutation operation of two vector fields $\alpha =\sum_{i=1}^nv_i\partial/\partial z_i$ and
$\beta =\sum_{i=1}^nu_i\partial/\partial z_i$.
In coordinate representation it has the standard form
\begin{equation}
[\alpha, \beta]_{W_n} \stackrel{{\rm def}}{=}
\sum_{i, j = 1}^n(v_i\frac{\partial u_j}{\partial z_i}
- u_i\frac{\partial v_j}{\partial z_i})
\frac{\partial}{\partial z_j}
\end{equation}
in which $\{u_i\}_{i=1}^n$ and $\{v_i\}_{i=1}^n$ denote formal power series in $z_1, ..., z_n$.
This topological algebra denoted by $W_n$ {\rm \cite{Fuks}}.
\footnote{
The algebra $W_1$ is simplest but important example. It has topological basis of the fields ${\bf e}_k$ and commutator in this basis which are given by the formulas
$
{\bf e}_k = z^{k+1}\partial/\partial z\,,
[{\bf e}_k, {\bf e}_\ell] = (\ell-k){\bf e}_{{k+\ell}}\,,
$
where $k= -1, 0, 1, 2, ...$.
} 
The algebra $W_n$ itself has no ideals, i.e. it is simple;
it may be interpreted as the algebra of $\infty$-jets of smooth vector fields on ${\mathbb R}^n$.
\end{remark} 
The subalgebra of the algebra $W_n$, consisting of the vector fields $\sum_iv_i \partial/{\partial z_i}$, for which
$\{v_i\}_{i=1}^n$ belong to the $(k+1)$-st power of the maximal ideal of the ring of formal power series, where 
$k=-1, 0, 1, 2, ...$ is denoted by ${L}_k$. It is clear that 
\begin{equation}
W_n = {L}_{-1}\supset {L}_{0}\supset 
{L}_{1}\supset 
\cdots\,,\,\,\,\,\,\,\,\,\, 
[{L}_{k},\,\, {L}_{\ell}]\subset 
{L}_{k+\ell}\,.
\end{equation} 
When $\ell > k\geq 0$ the algebra ${L}_{\ell}$ is an ideal in ${L}_{k}$. 
Vector fields $\sum_iv_i\partial/{\partial z_i}$ with trivial divergence $\sum_i\partial v_i/{\partial z_i}$ constitute the classical subalgebra $S_n$,
$
\parallel a_{ij}\parallel_{S_n}\longmapsto
\sum_{i=1}^n a_{i, n+1}\partial/\partial z_i,
$
of the algebra of formal vector fields $W_n$. This algebra is simple, but itself an ideal of codimension one in the algebra ${\widehat S}_n$, consisting of vector fields with constant divergence. 
The formula $\parallel a_{ij}\parallel\mapsto 
\sum a_{ij}z_i\partial/{\partial z_j}$ determines the canonical inclusion 
$
{\rm g}{\rm l}_n
\rightarrow {L}_{0} \subset W_n\,.
$
The composition of this inclusion yields the canonical isomorphism 
\begin{equation}
\xymatrix@=15mm{
{\rm g}{\rm l}_n
\ar@{->}[r]^{{\rm Inclusion}}
\ar@{->}[rd]_{\rm Isomorphism}
&{L}_{0}
\ar[d]^{{\rm Projection}}
\\
&{L}_{0}/{L}_{1}
}
\label{foot}
\end{equation}

\section{Invariants of truncated Weyl algebras}

Define a DG-algebra 
$\widetilde{W}^\bullet({\rm g}{\rm l}_n)$ as a factor algebra of the Weyl algebra $W^\bullet({\rm g}{\rm l}_n)$ (see Appendix \ref{Weyl}) over $(2n+1)$-term of standard filtration $F^\bullet$ in it. 
The following result holds: 
\begin{theorem}\, {\rm (D. B. Fuks \cite{Fuks}, Theorem $2.2.4^\prime$)}
There is the homomorphism
\begin{equation}
{\widetilde W}^{\bullet}({\rm g}{\rm l}_n) =
W^{\bullet}({\rm g}{\rm l}_n) / F^{2n+1} 
W^{\bullet}({\rm g}{\rm l}_n)\longrightarrow C^\bullet 
( W^{\bullet}({\rm g}{\rm l}_n)).
\label{HOM}
\end{equation}
The quotient complex ${\widetilde W}^{\bullet}({\rm g}{\rm l}_n)$
is usually known as {\it the truncated Weyl algebra}.
The homomorphism {\rm (\ref{HOM})} induces an isomorphism 
in cohomology. 
\end{theorem}
The proof of this statement uses a spectral sequence associated with filtration
\begin{equation}
\{F^\bullet W^{\bullet}({\rm g}{\rm l}_n)/F^{2n+1}
W^{\bullet}({\rm g}{\rm l}_n)\}
\,\,\,\,\, {\rm in} \,\,\,\,\, W^{\bullet}({\rm g}{\rm l}_n).
\label{HOM1}
\end{equation} 
Note that this latter spectral sequence satisfies Eq. (\ref{SS1}) of the Appendix A (see also Eq. (\ref{SS2})). The homomorphism (\ref{HOM}) is compatible with the filtration (\ref{HOM1}) and with the Hochschild-Serre filtration in complex $C^\bullet (W_n)$, so that we obtain a homomorphism of one spectral sequence to another. This homomorṕhism establishes an isomorphism between the terms $EW_2$ and even between the parts $EW_2^{0, q}, EW_2^{p, 0}$ of these terms.

\subsection{Tensor modules and the ${\rm g}{\rm l}_n$ invariants}
We concentrate on ${\rm g}{\rm l}_n$ tensor modules and it invariants. 
\footnote{
The main example of a ${\rm g}{\rm l}_n$ moduleis is the space ${ A}^n$ of column-vectors $V$ on which matricies act by left multiplication. Note also the one-dimensional module $E_\lambda$, where $\lambda\in {A}$. The structure of the module is determed by the formula $ga = -\lambda ({\rm Tr}\, g) a$, where ${\rm Tr}$ denotes trace. It is clear that $E_\lambda \otimes E_\mu = E_{\lambda+ \mu}$. Modules of the form $V\otimes \ldots \otimes V\otimes V^\prime\otimes\ldots\otimes V^\prime$ (and their submodules), where the prime denotes dual space are called tensor modules. Modules of the form $V\otimes\ldots\otimes V\otimes V^\prime\otimes\ldots\otimes V^\prime\otimes E_\lambda$ (and their submodules) are called generalized tensor modules.}
Denote by $V$ a tautological $n$-dimensional 
${\rm g}{\rm l}_n$-representation. 
Following \cite{Fuks} denote the elements of space $V$ by letters $\alpha$ while the elements of dual space $V^\prime$ by letters $\beta, \mu$. Then a tensor from the space 
$V^{\prime \otimes k}\otimes V^{\otimes \ell}$ is a function of variables $(\alpha_1, \cdots , \alpha_k; \beta_1, \cdots, \beta_\ell)$. 
\begin{theorem} \label{TM}
{\rm (The main theorem of invariants \cite{Weyl})}  
If $k\neq \ell$ then $\Inv \,T^k_\ell 
\equiv [V^{\prime \otimes k}\otimes 
V^{\otimes \ell}]^{{\rm g} {\rm l}_n}= 0$.
The space 
${\rm Inv}\,T_\ell^k$  with both
subscript $\ell$ and superscript $k=\ell$ is generated by tensors
$
c_\sigma = \sum_{i_1, \ldots, i_k =1}^n e_{\sigma(i_1)\cdots \sigma(i_k)}^{i_1\cdots i_k},
$
where $\sigma\in S_k\equiv {\rm Symm} (k)$.
\footnote{
Under the identification of $T_k^k$ with $(T_k^k)^\prime$ and with ${\rm Hom}(V\otimes \cdots \otimes V,\,V\otimes \cdots \otimes V )$, the tensor $c_\sigma$ becomes, respectively, the functional and the homomorphism described by the formulas
\begin{eqnarray}
\beta_1\otimes \cdots \otimes \beta_k\otimes \alpha_1\otimes \cdots \otimes \alpha_k & \longmapsto & \beta_1(\alpha_{\sigma(1)})
\cdots \beta_k(\alpha_{\sigma(k)})\,,
\nonumber \\
\alpha_1\otimes \cdots \otimes \alpha_k & \longmapsto &
\alpha_{\sigma(1)}\otimes \cdots \otimes \alpha_{\sigma(k)}\,.
\nonumber
\end{eqnarray}
}
The space  $[V^{\prime \otimes k}\otimes V^{\otimes k}]^{{\rm g}{\rm l}_n}$ is generated by traces, i.e. functionals of the kind 
$c_\sigma(\alpha_1, \cdots , \alpha_k; \beta_1, \cdots, \beta_k)=\beta_1(\alpha_{\sigma(1)})\cdot\ldots\cdot
\beta_k(\alpha_{\sigma(k)})$, where $\sigma$ is an element of symmetric group $S_k$.
If $k\leq n$, then the elements $c_\sigma$ are linear independent. If $k=n+1$ then there is the unique relation 
\begin{equation}
\sum_{\sigma\in S_{n+1}}{\rm sgn}(\sigma)c_\sigma= 0\,.
\label{relation}
\end{equation} 
If $k\geq n+1$ then all relations are algebraic consequences of the relation {\rm (\ref{relation})}. 
\end{theorem} 
Thus using the above notation one can consider the elements of space
\begin{equation}
\cT_{\overline{p}, q} = \bigotimes_{i\geq0, i\neq 1}
\Lambda^{p_i}(S^iV\otimes V^\prime)\otimes V^{\otimes q}\,,
\,\,\,\,\,\,\,\,\,
\overline{p} = (p_0, p_2, p_3, \ldots)
\label{B}
\end{equation}
as the linear function of set of variables:
\begin{eqnarray}
\Lambda^{p_0}V^\prime && \!\!{\rm corresponding}\,\,\, {\rm variables}\,\,\, {\rm are}:\,\,\, \alpha_1; \alpha_2; \ldots ;\alpha_{p_0}\,.
\nonumber \\
\Lambda^{p_2}(S^2V\otimes V^\prime) && \!\!{\rm corresponding}\,\,\, {\rm variables}\,\,\, {\rm are}:\,\,\, \beta^1_{p_0+1}, \beta^2_{p_0+1}, \alpha_{p_0+1}; \ldots ;\beta^1_{p_0+2}, \beta^2_{p_0+p_2}, \alpha_{p_0+p_2}\,.
\nonumber \\
\Lambda^{p_3}(S^3V\otimes V^\prime) && \!\!{\rm corresponding}\,\,\, {\rm variables}\,\,\, {\rm are}:\,\,\, \beta^1_{p_0+p_2+1}, \beta^2_{p_0+p_2+1}, \beta^3_{p_0+p_2+1},\alpha_{p_0+p_2+1}; \ldots ;
\nonumber \\
&&\!\!
\beta^1_{p_0+p_2+p_3}, \beta^2_{p_0+p_2+p_3}, \beta^3_{p_0+p_2+p_3}, \alpha_{p_0+p_2+p_3}\,.
\nonumber \\
{\cdots\cdots\cdots\cdots}
&& \cdots\cdots\cdots \cdots\cdots\cdots \cdots\cdots\cdots
\cdots\cdots\cdots \cdots\cdots\cdots,
\nonumber \\
V^{\otimes q} &&\!\! {\rm corresponding}\,\,\, {\rm variables}\,\,\, {\rm are}:\,\,\, \mu_1; \mu_2; \ldots ; \mu_q
\label{var}
\end{eqnarray}
The elements of the space $\cT_{\overline{p}, q}$ as function of variables $(\alpha, \beta, \mu)$ are subject to certain symmetry  conditions.
Namely, a tensor has to be symmetric with respect to variables $\beta$ with identical lower index, and it must change the sign under the following permutations:
\noindent
\\
\\
{\bf (i)}\,\, Permutation of two $\alpha$'s in the first line of (\ref{var})
$\alpha_i$ and $\alpha_j$\, $1\leq i\neq j\leq p_0$.
\noindent
\\
\\
{\bf (ii)}\,\, Permutation of group of variables of the second line of (\ref{var}) (which are separated by a punctuation mark "\,;\,"), i.e. the permutation $\beta^1_i, \beta^2_i, \alpha_i$ with the group
$\beta^1_j, \beta^2_j, \alpha_j$, where $p_0< i\neq j\leq p_0+p_2$. 
\noindent
\\
\\
{\bf (iii)}\,\, Permutation of similar groups in the third line of (\ref{var}): permutation of the group $\beta^1_i, \beta^2_i, \beta^3_i, \alpha_i$ with the group $\beta^1_j, \beta^2_j, \beta^3_j, \alpha_j$, where $p_0+p_2< i\neq j\leq p_0+p_2+p_3$; $\ldots$, etc.

\subsection{Lie algebras $W_n\ltimes {\overline G}\otimes P_n$}

\begin{lemma} \label{L1}
The relative cochain complex of the Lie algebra $W_n\ltimes {\overline G}\otimes {P}_n$ with respect to the Lie subalgebra 
${\rm g}{\rm l}_n$ coincides with the factor-algebra of the relative Weyl algebra over module of the same Lie subalgebra ${\rm g}{\rm l}_n$,
$
{\widetilde W}^\bullet({\rm g}{\rm l}_n\oplus {\overline G}, 
{\rm g}{\rm l}_n) =
W^\bullet({\rm g}{\rm l}_n\oplus {\overline G}, {\rm g}{\rm l}_n)
/F^{2n+1}W^\bullet({\rm g}{\rm l}_n\oplus {\overline G}, 
{\rm g}{\rm l}_n).
$ 
\end{lemma}
Let us analyze the structure of $k$-cochains of relative cochain complex. 
Recall that $q$-dimensional cochain of the algebra $\mathfrak g$ with coefficients in $A$ is a skew-symmetric $q$-linear functional on $\mathfrak g$ with values in $A$. The space of all such cochains is $C^q({\mathfrak g}; A)= {\rm Hom}(\Lambda^q{\mathfrak g}; A)$ and this representation transforms $C^q({\mathfrak g}; A)$ into $\mathfrak g$-module.
Since 
$
C^q(\mathfrak{g}, \mathfrak{h}; A) = {\rm Hom}_{\mathfrak{h}}(\Lambda (\mathfrak{g}/\mathfrak{h}); A), 
$
where $\mathfrak h$ is a subalgebra of the Lie algebra $\mathfrak g$ and $A$ is an arbitrary $\mathfrak h$-module, we have
\begin{equation}
C^k(W_n\ltimes{\overline G}
\otimes P_n, {\rm g}{\rm l}_n; A) =
\Hom_{{\rm g}{\rm l}_n}(\Lambda^k((W_n\ltimes{\overline G}
\otimes P_n)/{\rm g}{\rm l}_n); A)\,.
\end{equation}
Consider the structure of ${\rm g}{\rm l}_n$-representations in formal functions and vector fields. The algebra $W_n$ contains ${\rm gl}_n$ as a subalgebra, and therefore it is a ${\rm gl}_n$-module. For every $k$ the vector fields $\sum F_k\partial/\partial z_k$ constitute a submodule of this module. It is clear that this submodule is isomorphic to $S^kV^\prime\otimes V$. 
Thus we have
\begin{equation}
{\rm g}{\rm l}_n \cong V^\prime\otimes V,\,\,\,\,\,\,\,\,\,\,
P_n \cong \widehat{\bigoplus}_{0\leq m < \infty} S^mV^\prime,\,\,\,\,\,\,\,\,\,
W_n \cong  \widehat{\bigoplus}_{0\leq m < \infty} S^mV^\prime
\otimes V,
\end{equation}
where the notation ${\widehat{\bigoplus}}$ means a direct sum enlarged with respect to 
the formal power series. Recall that $SV = \oplus S^jV$ denotes the symmetric algebra of a $A$-module. In the case of extended Lie algebra 
$W_n\ltimes{\overline G}\otimes P_n$ we have
\begin{eqnarray}
((W_n\ltimes{\overline G}\otimes P_n)/{\rm g}{\rm l}_n)^*
& \cong & 
\left(
(\widehat{\bigoplus}_{i>0, i\neq 1} S^iV^\prime\otimes V)
\oplus {\overline G}\otimes (\widehat{\bigoplus}_{j\geq 0}
S^jV^\prime )\right)^*
\nonumber \\
& \cong &
(\bigoplus_{i>0, i\neq 1} S^iV\otimes V^\prime)\oplus {\overline G}^*
\otimes (\bigoplus_{j\geq 0}S^jV)\,.
\end{eqnarray}
The space of $k^{th}$ cochains can be represented as a sum of invariants in tensor products 
\begin{eqnarray}
&& C^k(W_n\ltimes{\overline G}
\otimes P_n, {\rm g}{\rm l}_n; A)  =  
\left[\Lambda^k\left(
(\bigoplus_{i>0, i\neq 1} S^iV\otimes V^\prime)\oplus {\overline G}^*
\otimes (\bigoplus_{j\geq 0}S^jV)\right)\right]^{{\rm g}{\rm l}_n}
\nonumber \\
&& \,\,\,\, = 
\left[\bigoplus_{\atop{\scriptstyle p_0+p_2+p_3+ \cdots
\atop{\scriptstyle 
+q_0+q_1+q_2\cdots = k}}}
\left(\begin{array}{ll}
\Lambda^{p_0}V^\prime\otimes \Lambda^{p_2}(S^2V\Lambda V^\prime)\otimes \Lambda^{p_3}(S^3V\otimes V^\prime)\otimes \cdots\\
\otimes \Lambda^{q_0}{\overline G}^*\otimes \Lambda^{q_1}
({\overline G}^*V)
\otimes \Lambda^{q_2}({\overline G}^*\otimes S^2V)\otimes\cdots
\end{array} \right)
\right]^{{\rm g}{\rm l}_n}\!.
\end{eqnarray}
The ring of invariants $[\cT_{\overline{p}, q}]^{{\rm g}{\rm l}_n}$ (Eq. (\ref{B})) is
generated by trace $t\in [V^\prime\otimes V]^{{\rm g}{\rm l}_n}$, determined by $t(\alpha; \mu) = \mu(\alpha)$, and elements 
$\Psi_r\in [\Lambda^rV^\prime\otimes \Lambda^r(S^2V\otimes V^\prime)]^
{{\rm g}{\rm l}_n}$. The elements $\Psi_r$ are given by
\begin{eqnarray}
&&\Psi_r(\alpha_1; \cdot\cdot\cdot ; \alpha_r; \beta_{r=1}^2, \alpha_{r+1}; \cdot\cdot\cdot ; \beta_{2r}^1, \beta_{2r}^2, \alpha_{2r})
\nonumber \\
&&
=
\sum_{\atop{\scriptstyle \sigma, \tau \in S_r
\atop{\scriptstyle 
\nu_1, \cdot\cdot\cdot , \nu_r\in S_2}}}
\left[ {\rm sgn}(\sigma){\rm sgn}(\tau)\prod_{j=1}^r \beta_{r+\tau(j)}^{\nu_j(1)}(\alpha_{\sigma(j)})
\beta_{r+\tau(j)}^{\nu_j(2)}(\alpha_{r+\tau(j-1)})\right]\,,
\end{eqnarray}
where we assume that $\tau(0)=\tau(r)$.
We refer the reader to the paper \cite{Khoroshkin} where 
${\rm g}{\rm l}_n$-invariants involve linear and constant 
$\overline G$-valued functions. Here we restrict ourselves to the final result. Relative cochains complex has the form:
\begin{eqnarray}
\!\!\!\!\!\!\!\!\!\!\!\!\!\!\!
C^k(W_n\ltimes{\overline G}\otimes P_n, {\rm g} {\rm l}_n; A)  & = & \left[\Lambda^k((W_n\ltimes {\overline G}\otimes {P}_n, {\rm g} {\rm l}_n; {A})/{{\rm g} {\rm l}_n})^*\right]^{{\rm g} {\rm l}_n} 
\nonumber \\
& = & 
\bigoplus_{m+\ell = k}\left[\Lambda^m(W_n/{{\rm g} {\rm l}_n})^*
\otimes \Lambda^\ell ({\overline G}^*\otimes P_n)^*\right]^{{\rm g} {\rm l}_n}
\nonumber \\
& = & 
\!\!\!\!\!\!
\bigoplus_{2p_0+q_0+2q_1=k}\!\!\!\left[
\Lambda^{p_0}V^\prime \otimes \Lambda^{p_0+q_1}(S^2V\otimes V^\prime)
\otimes \Lambda^{q_1}V\right]
\otimes \cC({\overline G}; q),
\end{eqnarray}
where  $\cC({\overline G}; q) = \Lambda^{q_0}{\overline G}^*\otimes S^{q_1}{\overline G}^*$ and
\begin{equation}
\bigoplus_{k\geq 0}C^k(W_n\ltimes{\overline G}\otimes P_n, {\rm g} {\rm l}_n; A) = \langle t^{q_1}\Psi_1^{r_1}\Psi_2^{r_2}
\cdot\cdot\cdot  \Psi_n^{r_n}\vert q_1+r_1+2r_2 + \cdot\cdot\cdot + nr_n\leq n \rangle
\otimes (\bigoplus_{q_j}\cC({\overline G}; q)).
\end{equation}
It can be checked that differentials in the cochain complex and in the Weyl factor-algebra $W^\bullet ({\rm g}{\rm l}_n\oplus {\overline G}, {{\rm g}{\rm l}_n})$ are identical; this completes the proof of Lemma \ref{L1}.

\section{Morphisms of filtered complexes}
The algebra $\widetilde{W}^\bullet({\rm g}{\rm l}_n\oplus {\overline G})$ has filtration from Weyl algebra. The following main 
statement asserts the structure of a cohomology ring of the Lie algebra ${\rm g}{\rm l}_n\oplus {\overline G}$:  
The filtered DG-algebra $F^\bullet\widetilde{W}^\bullet({\rm g}{\rm l}_n\oplus {\overline G})$ is quasiconformal to cochain complex 
of Lie algebra $W_n\ltimes {\overline G}\otimes {P}_n$ with constant coefficients and Hochshild-Serre filtration with respect to Lie 
subalgebra ${\rm g}{\rm l}_n\oplus {\overline G}$. 

In terms of spectral sequences this statement can be reformulated as follows. 
Let ${\mathfrak h}$ be a subalgebra of a Lie algebra ${\mathfrak g}$. Choose an ${\mathfrak h}$-module direct sum decomposition of 
${\mathfrak g}, \, {\mathfrak g} = {\mathfrak h}\oplus {\mathfrak g}/{\mathfrak h}$, and let ${\pi}$ denote both the corresponding 
projection of ${\mathfrak g}$ onto ${\mathfrak h}$ and onto ${\mathfrak g}/{\mathfrak h}$.
By analogy with Leray-Serre filtration (Eqs. (\ref{Filt1}), (\ref{Filt2})), one can introduce filtration $\Phi^\bullet$ on a cochain 
complex of Lie algebra ${\mathfrak g}$. 
The projection $\pi$ assigns morphism of filtered DG-algebras
\begin{equation}
\overline{\pi}: F^\bullet W^\bullet ({\mathfrak h}) \longrightarrow \Phi^\bullet C^\bullet ({\mathfrak g})\,.
\label{pi1}
\end{equation}
There is a morphism of corresponding spectral sequences and in particular their first terms are:
\begin{equation}
{\pi}^*: H^q({\mathfrak h}; S^p{\mathfrak h}^*)\longrightarrow
H^q({\mathfrak h}; \Lambda^{2p}({\mathfrak g}/{\mathfrak h})^*)\,.
\end{equation}
If ${\mathfrak h}, {\mathfrak g}$ are the Lie algebras of compact Lie groups $H, G$, then such a morphism is a
particular case of the Chern-Weyl homomrphism for $H$-fibering 
$G \stackrel{H}{\rightarrow} G/H$, with the following difference: the deRham complexes have to be substituted for cohomological 
complexes of left invariant forms. Now let us ascertain an expansion of 
$({\rm g}{\rm l}_n\oplus {\overline G})$-modules
\begin{equation}
\alpha:\, W_n\ltimes {\overline G}\otimes {P}_n \stackrel{\sim}
{\longrightarrow}({\rm g}{\rm l}_n\oplus {\overline G})\oplus
((W_n\ltimes {\overline G}\otimes {P}_n)/
({\rm g}{\rm l}_n\oplus {\overline G}))\,.
\label{alpha}
\end{equation}
Consider a diagram of morphisms of filtered complexes \cite{Khoroshkin}
\begin{equation}
\xymatrix@C=10mm{
&F^\bullet W^\bullet({\rm g}{\rm l}_n\oplus {\overline G})
\ar[dl]\ar[dr]^{\overline{\alpha}}&\\
F^\bullet {\widetilde W}^\bullet({\rm g}{\rm l}_n
\oplus {\overline G})&&
\Phi^\bullet C^\bullet (W_n\ltimes {\overline G}\otimes {P}_n;
A) \\
}
\label{D}
\end{equation}
Here $\overline \alpha$ has been constructed from $\alpha$ as it has been discussed above (formula (\ref{pi1})). The corresponding 
diagram of spectral sequences takes the form
\begin{equation}
\xymatrix@C=10mm{
&E^\bullet W^{\bullet, \bullet}({\rm g}{\rm l}_n\oplus 
{\overline G})
\ar[dl]\ar[dr]^{\overline{\alpha}}&\\
E{\widetilde W}^{\bullet, \bullet}({\rm g}{\rm l}_n\oplus 
{\overline G})&&
EC^{\bullet, \bullet} (W_n\ltimes {\overline G}\otimes {P}_n;
A) \\
}
\label{D1}
\end{equation}
Spectral sequences in the lower line of diagram {\rm (\ref{D1})}
coincide beginning from the first term.

\section{BRST cohomology, characteristic classes of foliations and characters}

The sheaf cohomology group $H^p(X,\Lambda^qTX)$ associated with $(0,q)$ forms on space 
$X$ with values in $\Lambda^qTX$ (the $q^{th}$ exterior power of the holomorphic tangent 
bundle of $X$), consists of solutions of $\overline{\partial}{\mathcal V}=0$ modulo 
${\mathcal V}\rightarrow {\mathcal V} + \overline{\partial} S$ for an object
\begin{eqnarray}
{{\cO}}_{\mathcal V} & = & \eta^{\bar i_1}\eta^{\bar i_2}\dots \eta^{\bar i_p}
{\mathcal V}_{{\bar i_1}{\bar i_2}\dots \bar i_p}{}^{j_1 j_2\dots j_q}\psi_{j_1}
\dots \psi_{j_q}
\nonumber \\
& = &
d\bar z^{{\bar i}_1}d\bar z^{{\bar i}_2}\dots d\bar z^{{\bar i}_p}{\mathcal V}_{\bar i_1\,
\bar i_2\dots \bar i_p}{}^{j_1j_2\dots j_q}{\partial/\partial z_{j_1}}
\dots {\partial/\partial z_{j_q}}\,,
\label{operator}
\end{eqnarray}
which is called the vertex operator.

Recall that for the topological field theory the transformation laws in terms of BRST 
operator $Q$ can be expressed as $\delta W = -\{Q, W\}$
for any field $W$; in addition $Q^2 = 0$. Standard arguments using the $Q$ invariance 
show that the correlation function $\langle \{Q, W\}\rangle = 0$ for any $W$. It is also 
true that $\{Q, {\cO}_a\} = 0$ for some BRST invariant operators ${\cO}_a$. 
As before (see Eq. (\ref{QW})) the local variables are
$
\eta^{\bar i} \in {\Gamma}(\phi^*T^{\vee}X) 
\sim d{\bar z}^{\bar i}\,,\,\,\,\,\,\,\,
\theta_i \in g_{i{\bar j}}({\Gamma}(\phi^*T^{\vee}X))^{\bar j}
\sim \partial/{\partial z^i}\,.
$
Mathematically one can interpret $\eta^{\overline{i}}$ as the (0, 1) form 
$d\phi^{\overline{i}}$, and $\phi^{\overline{i}} \sim \overline{z}^{\overline{i}},\,
\eta^{\overline{i}}\sim d\phi^{\overline{i}}$.
The Noether BRST charge $Q$ acts on the fields as 
$
Q\sim \overline{\partial}  =  \sum_i\eta^{\overline i}\partial/\partial\phi^{\overline i}
\sim \sum_id{\overline z}^{\overline i}
\partial/\partial {\overline {z}}^{\overline i}\,.
$
Then
\begin{equation}
\delta_Q(\cO_{\mathcal V}) = {\cO}_{\overline{\partial} {\mathcal V}};\,\,\,\,\, 
{\cO}_{\mathcal V}\,\,\, {\rm is}\,\,\, {\rm BRST}\,\,\, 
{\rm invariant}\,\,\, {\rm if}\,\,\, {\overline{\partial}}{\mathcal V}=0
\,\,\, {\rm and}\,\,\, {\rm BRST}\,\,\, {\rm exact}\,\,\, 
{\rm if}\,\,\, {\mathcal V} = \overline{\partial} S\,.
\label{vertex}
\end{equation}
The BRST transformation (for bosonic and fermionic fields) is nilpotent, $\delta_Q^2 = 0$, 
and is a derivation of the algebra generated by fields and their derivatives. 
There is a natural map ${\mathcal V}\rightarrow {\cO}_{\mathcal V}$ from 
$\oplus_{p,q}H^p(X,\Lambda^q TX)$ 
to the BRST cohomology (see also Section \ref{LAH}). Therefore the BRST cohomology is 
isomorphic to the Dolbeault cohomology, and {\it on the classical level} the BRST operator 
acts as Dolbeault operator. The explanation can be found in a book by Nakanishi
and Ojima \cite{Nakanishi} about the mutual relations among gauge
and BRST transformations, BRST cohomology with its implementing BRST operator (or BRST charge) 
and their roles in determining the physical contents of the theory.

A relevant concept in the vertex operator algebra and the BRST cohomology is that of elliptic genus. Elliptic genera are natural topological invariants, which generalize the classical genera. They appear when one considers supersymmetric indices of the superconformal vertex algebras (SCVA). For mathematicians, elliptic genera (and the respective elliptic cohomology) may be associated 
to new mathematical invariants for spaces, while for physicists elliptic genera are the one-loop string partition functions. In many applications, when the space $X$ is a Calabi-Yau manifold, there is an $N = 2$ SCFT associated to $X$, with the two twists leading to the A-model and B-model, respectively. This has been discussed over the years in an impressive number of papers. The mathematical formalization of these concepts has also existed for some time. A sheaf of the topological vertex algebras for any Calabi-Yau manifold has been constructed in \cite{Malikov}, where the theory involved is the A-model. A different approach based on standard techniques in differential geometry has been developed in \cite{Zhou00}, where the holomorphic vector bundles of the $N = 2$ SCVA on a complex manifold $X$ and the $\overline{\partial}$ operator on such bundles have been used.

For a holomorphic vector bundle $E$ on $X$ and a formal variable $z$ we use the following 
identities
\begin{eqnarray}
S_q \left( z{E} \right) & = & 1 \: \oplus \: z q {E} \: \oplus \:
z^2 q^2 \mbox{Sym}^2 {E} \: \oplus \:
z^3 q^3 \mbox{Sym}^3 {E} \: \oplus \: \cdots = S_{zq} {E}\,,
\\
\Lambda_q \left( z {E} \right) & = & 1 \: \oplus \: z q {E}
\: \oplus \:
z^2 q^2 \mbox{Alt}^2 {E} \,\,\,\: \oplus \:
z^3 q^3 \mbox{Alt}^3 {E}\,\,\, \: \oplus \: \cdots =  \Lambda_{zq} {E}\,,
\\
S_q \left( z {E} \right)^{ {\mathbb C} } & = &
S_q \left( z {E}\right) \otimes S_q \left(
\overline{z} \overline{ {E} } \right),\,\,\,\,\,\,\,
\Lambda_q \left( z {E}\right)^{ {\mathbb C} } =
\Lambda_q \left( z{E} \right) \otimes \Lambda_q
\left( \overline{z} \overline{ {E} } \right)\,.
\end{eqnarray}
These identities have good multiplicative properties and its elements should be understood 
as elements of the $K$-theory of the underlying space.
\begin{eqnarray}
S_q \left( {E} \oplus {F} \right)  & = &
\left( S_q {E} \right) \otimes \left( S_q {F} \right), \,\,\,\,\,\,\,
S_q \left( {E} \ominus {F} \right) =
\left( S_q {E} \right) \otimes \left(
S_q {F} \right)^{-1}\,,
\label{ident1}
\\
\Lambda_q \left( {E} \oplus {F} \right) & = &
\left( \Lambda_q {E} \right) \otimes
\left( \Lambda_q {F} \right), \,\,\,\,\,\,\,
\Lambda_q \left( {E} \ominus {F} \right) =
\left( \Lambda_q {E} \right) \otimes
\left( \Lambda_q {F} \right)^{-1}\,.
\label{ident2}
\end{eqnarray}
In Eqs. (\ref{ident1}), (\ref{ident2}) we have used the fact that
\begin{eqnarray}
\mbox{Sym}^n ({E} \oplus {F}) & = &
\bigoplus_{i=0}^n \, \mbox{Sym}^i({E}) \otimes
\mbox{Sym}^{n-i}({F})\,, \\
\mbox{Alt}^n ({E} \oplus {F}) & = &
\bigoplus_{i=0}^n \, \mbox{Alt}^i({E}) \otimes
\mbox{Alt}^{n-i}({F})\,.
\end{eqnarray}
In the case of a line bundle ${\mathcal L}$, we have
$
S_q {\mathcal L} = 1 \bigoplus_{n\in {\mathbb Z}_+} q^n {\mathcal L}^n
= (1 \ominus q {\mathcal L})^{-1} = ( \Lambda_{-q} {\mathcal L})^{-1},
$
and therefore
$\left( S_q {E} \right)^{-1} = \Lambda_{-q} {E}$
for any vector bundle ${E}$, and similarly
$\left( \Lambda_q {E} \right)^{-1} = S_{-q} {E}$.

Let us note some well known examples of the vertex operator algebra bundles which have been used in the literature to study the elliptic genus and the Witten genus.
If $X$ is a Riemannian manifold, then the transition functions of the complex tangent bundle $T_{\mathbb C}X$ lie in the special orthogonal group $SO(d)$, where $d$ is the dimension of $X$. Then $\bigotimes_{n\in {\mathbb Z}_+}S_{q^n}(T_{\mathbb C} X)$ is a $V_H^{SO(d)}$-bundle. Here $V_H$ is the Heisenberg vertex operator algebra of dimension $d$, with $SO(d)$ as a subgroup of Aut$({V_H})$, and ${V_H}^{SO(d)}$ is the set of $SO(d)$-invariants of $V_H$, which is the vertex operator subalgebra of $V_H$. Similarly, $\bigotimes_{n\in {\mathbb Z}_+\cup \{0\}} \Lambda_{q^{n+1/2}}(T_{\mathbb C} X)$ is an $L(1,0)^{SO(d)}$-bundle where $L(1,0)$ is the level one module for the affine algebra $D_{d/2}^{(1)}$. In this case we assume that $d$ is even.
If $X$ is further assumed to be a spin manifold, we denote the spin bundle by $\cS$. Then
$\cS\otimes\bigotimes_{n\in {\mathbb Z}_+}\Lambda_{q^{n}}(T_{\mathbb C} X)$ is also a $L(1,0)^{SO(d)}$-bundle.

The prototypes for the elliptic genera are the expressions (see for detail \cite{BBE}):
\begin{eqnarray}
&& 
\bigotimes_{n \in {\mathbb Z}_+}S_{\sigma q^n}((\xi\cP)^{\mathbb C})
\bigotimes_{n\in {\mathbb Z}_+}
\Lambda_{\lambda q^n}\left((\zeta{\cQ})^{\mathbb C}\right),
\,\,\,\,\,\,\,\,\,\,\,\,
\bigotimes_{n \in {\mathbb Z}_+}S_{\sigma q^n}((\xi\cP)^{\mathbb C})
\bigotimes_{n\in {\mathbb Z}_+/2}
\Lambda_{\lambda q^n}\left((\zeta{\cQ})^{\mathbb C}\right),
\\
&&
\!\!
\bigotimes_{n\in {\mathbb Z}_+/2}\!\!S_{\sigma q^n}((\xi\cP)^{\mathbb C})
\bigotimes_{n\in {\mathbb Z}_+}
\Lambda_{\lambda q^n}\left((\zeta{\cQ})^{\mathbb C}\right),
\,\,\,\,\,\,\,\,\,\,\,\,
\!\!\bigotimes_{n\in {\mathbb Z}_+/2}\!\!S_{\sigma q^n}((\xi\cP)^{\mathbb C})
\bigotimes_{n\in {\mathbb Z}_+/2}
\Lambda_{\lambda q^n}\left((\zeta{\cQ})^{\mathbb C}\right).
\end{eqnarray}

{\bf Characteristic classes of foliations.}
As before let us suppose that $\frak g$ is a Lie algebra (in the following we shall assume that $\frak g$ is a subalgebra of the Lie algebra $W_n$). By a $\frak g$-{\it structure} on a smooth manifold $X$ we mean, following the lines of the Bernstein-Rosenfeld article
\cite{B-R}, a smooth one-form $\omega$ on $X$ with values in $\frak g$, satisfying the Maurier-Cartan equation $d\omega = -(1/2)[\omega, \omega]$. The latter means that for any vector fields $\xi_1, \xi_2$ on $X$, $d\omega(\xi_1, \xi_2) = - [\omega(\xi_1), \omega(\xi_2)]$. 
As before, let us consider the pair $(H, G)$ of Lie groups. It follow that the pair $(H, G)$ 
with a discrete quotient group $({\rm Norm} H)/H)$ corresponds the inclusion ${\frak g}\rightarrow W_n$, where 
$n = {\rm dim} \,G/H$, while the quotient space $G/H$ possesses a canonical $\frak g$-structure (see for detail \cite{Fuks}). Combining this $\frak g$-structure and the above inclusion one can obtain a $W_n$-structure on $G/\Gamma$, where $\Gamma$ is a discrete subgroup of the group $G$, and this is precisely the $W_n$-structure wich corresponds to the foliation ${\frak F}(G, H, \Gamma)$.
Thus the homomorphism
\begin{equation}
{\rm Char :} \,\,\,\, H^\bullet (W_n)\longrightarrow H^\bullet (G/\Gamma; {\mathbb R})
\label{Char}
\end{equation}
splits up into the composition of homomorphisms
\begin{equation}
H^\bullet (W_n)\longrightarrow H^\bullet ({\frak g})\,,\,\,\,\,\,\,\,\,\,
H^\bullet ({\frak g}) \longrightarrow H^\bullet (G/\Gamma; {\mathbb R})
\label{Char1}
\end{equation}
of which the first has nothing to do with $\Gamma$ and is induced by the inclusion 
${\frak g}\rightarrow W_n$, while the second has nothing to do with $H$ and corresponds to the canonical $\frak g$-structure on $G/\Gamma$.

For the canonical $\frak g$-{\it structure} on $G/\Gamma$ characteristic classes are determined by the canonical homomorphism
$
H^\bullet({\frak g}) \longrightarrow H^\bullet(G/\Gamma; {\mathbb R}). 
$
Suppose that the algebra $\frak g$ is {\it unitary} (i.e. $H_n({\frak g}) \neq 0$)
\footnote{If the algebra $\frak g$ is determined by means of the structural 
constants $c_{ij}^k$, $[e_i, e_j] = \sum_{k=1}^nc_{ij^ke_k}$ for some basis 
$e_1, \cdots, e_n$ of the space $\frak g$, then the unitary condition can be written in the form 
$\sum_{j=1}^nc_{ij^j} =0, i=1,..., n$. Note that semisimple and nilpotent Lie algebras are 
unitary algebras.} 
and the quotient space $G/\Gamma$ is compact, then this homomorphism is a monomorphism 
\cite{Fuks}. If the group $G$ is semisimple, then the algebra $\frak g$ is unitary and $G$ has a discrete subgroup $\Gamma$ with compact $G/\Gamma$. If the group $G$ is semisimple, then for an appropriately choosen group $\Gamma$ the kernel of the homomorphism 
$H^\bullet (W_n)\longrightarrow H^\bullet (G/\Gamma; {\mathbb R})$
coincides with the kernel of the homomorphism $H^\bullet (W_n)\rightarrow H^\bullet ({\frak g})$. In general the second kernel in Eq. (\ref{Char1}) is contained in the first one.

{\bf The Lefschetz formula and the Chern polynomials.}
The purpose of this section is to establish the Lefschetz fixed point formula
(which counts the number of fixed points of a continuous mapping from a compact
topological space to itself) with its connection to the Chern polynomials. 
Recall that the vertex operator algebras can be constructed from the highest weight representations of infinite dimensional Lie algebras. The characters of (integrable) highest weight modules can be identified with the holomorphic parts of the partition functions (elliptic genera) on the torus for the corresponding conformal field theories. All these structures arise naturally, but not exclusively, in string theory, and are particularly clear and treatable when supersymmetry 
is involved.

Let $X$ be a compact complex manifold and let $G$ be a Lie group acting on $X$ by biholomorphic maps. Let $g$ be a generator of $G$.
Furthermore, let $\pi: E \rightarrow X$ be a holomorphic vector bundle which admits a $G$-action compatible with the $G$-action on $X$. Let $G$ be a compact Lie group, then a characteristic class $G$ can be defined as a functor which assigns to every principal $G$-bundle $P$
a cohomology class of $X = P/G$. The set of all characteristic classes forms a ring $H^*_G(A)$ where the coefficient ring $A$ has to be specify. 
For example, let $G= T$ be a torus, and $T^*$ a character group (or Pontryagin dual) of $T$. 
Let $\{x_i\}_{i=1}^n$ are basis for $T^*$, then $H^*_G(A) = A[[x_1, \cdots x_n]]$ is the ring 
for formal power series in $x_1, \cdots, x_n$.

For the complex vector bundle $N^g$ one can construct a decomposition
$N^g = \sum N^g(\theta)$, where $N^g(\theta)$ is the sub-bundle on which $g$ acts as $\exp(i\theta)$, and \cite{Atiyah68}
\begin{equation}
{\rm ch}\,\Lambda_{-1}(N^g(\theta))^* = \prod_j (1-e^{-x_j-i\theta})
(N^g(\theta))\,.
\end{equation}
Here $\prod_j (1- e^{-x_j-i\theta}) \in H^*_{U(m)}({\mathbb C})$,
$m={\rm dim}\,N^g(\theta)$. For $0 <\theta< 2\pi$ define the stable characteristic class
$
{\mathfrak U}^{\theta} = \sum {\mathfrak U}^{\theta}_r =
\prod_j\left[\frac{1- e^{-x_j-i\theta}}{1- e^{-i\theta}}\right]^{-1}.
$
Thus each $\mathfrak U^{\theta}_r$ is a polynomial with complex coefficients in the Chern classes, and
\begin{equation}
[{\rm ch}\,\Lambda_{-1}(N^g(\theta))^*]^{-1} =
\frac{\mathfrak U^{\theta}(N^g(\theta))}{(1- e^{-i\theta})^m}.
\end{equation}
Taking the product over all $\theta$, we get
\begin{equation}
[{\rm ch}\,\Lambda_{-1}(N^g)^*]^{-1} =
\frac{\prod\,{\mathfrak U}^{\theta}(N^g(\theta))}{{\rm det}\,(1- g(N^g)^*)},
\end{equation}
where ${\rm det}_{\mathbb C}(1 - g | (N^g)^*)\in H^0(X^g; {\mathbb C})$
assigns to the component of $x\in X^g$ the value ${\rm det}_{\mathbb C}(1 - g | (N^g_x)^*)$.

Finally, let $X$ be a compact complex manifold and $E$ a holomorphic vector bundle over X. Suppose that $G$ is a finite group of automorphisms of the pair $(X, E)$. For any $g\in G$, let $X^g$ denote the fixed point set of $g$, and let, as before, $N^g = \sum N^g(\theta)$ denote the (complex) normal bundle of $X^g$ decomposed according to the eigenvalues $\exp (i\theta)$ of $g$. Let ${\mathfrak U}^{\theta}$ denote the characteristic class. Then (combinning the Lefschetz theorem with the Riemann-Roch theorem) one gets \cite{Atiyah68}
\begin{eqnarray}
\sum (-1)^p {\rm Tr}\,(g \,|\, H^p(X; \cO(E)) & = &
\left\{\frac{{\rm ch}\,(E | X^g)(g){\rm Td}(X^g)}{{\rm ch}\,\Lambda_{-1}((N^g)^*)(g)}\right\}[X^g]
\label{AS1}
\nonumber \\
& = &
\left\{\frac{{\rm ch}\,(E | X^g)(g)\, \prod_{\theta}{\mathfrak U}^{\theta}(N^g(\theta)){\rm Td}(X^g)}{{\rm det}\,(1- g | (N^g)^*)}\right\}[X^g]\,.
\label{AS2}
\end{eqnarray}
In Eq. (\ref{AS2}) ${\rm Td}(X^g)$ is the {\it Todd class}. Formally, the Todd class and the 
{\it dual Todd class} are ${\rm Td} = \prod_{j=1}^n(x_j/(1-e^{-x_j}))$
and ${\rm Td}^* = \prod_{j=1}^n(-x_j/(1-e^{x_j}))$ respectively. 
If $E^*$ is the dual to ${E}$ then 
${\rm Td}({E})$ = ${\rm Td}({E}^*)$. In particular, for the complexification of a real bundle,
${E} = E\bigotimes_{\mathbb R}{\mathbb C}$, ${E}\cong {E}^*$, and thus 
${\rm Td}({E}) = {\rm Td}^*({E})$.
The functor $E \mapsto {\rm Td}(E\bigotimes_{\mathbb R}{\mathbb C})$
defines a characteristic class of $O(n)$, and the image of $\rm  Td$ in the homomorphism $H^*_{U(n)}({\mathbb Q})\rightarrow H^*_{O(n)}({\mathbb Q})$. This class is called 
{\it the index class}. If $y_1, \cdots, y_m$ are the basic characters for the maximal torus of $O(n)$ ($m = [n/2]$), then
$
{\rm Td}(E) = {\rm Td}(E\bigotimes_{\mathbb R}{\mathbb C})
$, where
$
{\rm Td} = \prod (-y_j/(1- e^{y_j}))\prod (y_j/(1-e^{-y_j}).
$

In the Todd class ${\rm Td}$ the formal Chern roots $\{x_j\}_{j=1}^n$ of $TX$ are defined by ${\rm c} (TX) = \prod_{j} (1+x_{j})$. Then for $\bigotimes _{n\in {\mathbb Z}_+}
S_{q^n}((TX)^{\mathbb C})$ the resulting Chern character takes the form
\begin{eqnarray}
{\rm ch}(\bigotimes_{n\in {\mathbb Z}_+} S_{q^{n}}
((TX)^{\mathbb C})) 
& = &
\prod_j\prod_{n\in {\mathbb Z}_+}[(1-q^n e^{x_j})(1-q^n e^{-x_j})]^{-1}
\nonumber \\
& = &
\prod_j[\cR(s= \xi_j(1-it))\cdot \cR(s= -\xi_j(1-it))]^{-1}\,,
\end{eqnarray}
where $q = \exp (2\pi i\tau), t= {\rm Re} \tau/{\rm Im \tau}$ and $ \xi_j = x_j/2\pi i$.
\begin{remark}
One of the important features of the theory of infinite dimensional Lie algebras is the modular properties of characters of certain representations. The Chern polynomials {\rm (}and elliptic genera{\rm )} can be converted into product expressions which inherits modular and cohomology properties {\rm (}in sence of characteristic classes, Eq. {\rm (\ref{Char1}))}{\rm )} of appropriate {\rm (}polygraded{\rm )} Lie algebras. The final result can be written in terms of spectral functions of the hyperbolic three-geometry associated with $q$-series. The spectral Patterson-Selberg and Ruelle functions, $Z_\Gamma(s)$ and $\cR(s)$ respectively, can be attached to a closed oriented hyperbolic three-manifolds $X= H^3/\Gamma$ {\rm (}with acyclic orthogonal representation of $\pi_1(X)${\rm )} as follows 
{\rm \cite{Perry,BBE}}:
\begin{eqnarray}
&&
\!\!\!\!\!\!\!\!\!\!\!\!\!\!\!\!\!\!
Z_\Gamma(s) :=\!\!\! \prod_{k_1, k_2 \in \mathbb{Z}_+ \cup \{0\}} \!\!\![1-(e^{i\beta})^{k_1}(e^{-i\beta})^{k_2}e^{-(k_1+k_2+s)\alpha}],\,\,\,\,\,\,\,\,\,\,\,\,\,
{\mathcal R}(s)=\prod_{p=0}^{{\rm dim}\,X-1}Z_{\Gamma}
(p+s)^{(-1)^p},
\\
&&
\!\!\!\!\!\!\!\!\!\!\!\!\!\!\!\!\!\!
\prod_{n=\ell}^{\infty}(1- q^{n+\varepsilon})
= \cR(s=\xi(1-it))\,,\,\,\,\,\, \xi = \ell + \varepsilon.
\end{eqnarray}
In many applications the quantum generating functions of the topological field theories can be reproduced in terms of spectral functions of Selberg type. Therefore, the role of the unimodular group $SL(2;{\mathbb C})$ and of the modular group $SL(2; {\mathbb Z})$ constitute a clear manifestation of the remarkable link that exists between all the above and hyperbolic three-geometry.
\end{remark}

\section{The Hochschild cochain complex and deformations} \label{Hochschild}

\subsection{Lie algebra cohomology}
Let ${\rm gl}_n(A)$ be the Lie algebra of $n\times n$ matrices with coefficients in $A$. 
\footnote{
Cohomology of Lie algebra ${\rm gl}_n (A)$ with coefficients in finite representation $A$ coinsides with cohomology ${\rm gl}_n$ with coefficients in invariants of actions of ${\rm g}{\rm l}_n$ on $A$ \cite{Fuks},
$
H^\bullet_{\rm Lie}({\rm gl}_n; A)\cong
H^\bullet_{\rm Lie}({\rm gl}_n)
\otimes [A]^{{\rm gl}_n}\,.
$
}
In fact it is the tensor product $M_n({\mathbb C})\otimes {A}$ of the algebra of $n\times n$ matrices with $A$, considered as a Lie algebra. It contains the Lie subalgebra $M_n({\mathbb C})\otimes {\id}$.
Define the cochain $\Phi^{n} \in C^{k}({\rm gl}_n)$ by the formula 
\begin{equation}
\Phi^n(g_1, \ldots, g_{k}) \stackrel{{\rm def}}{=} 
\sum_{\sigma\in S_{k}}
\mathrm{sign}(\sigma)
\mathrm{Tr}(g_{\sigma(1)} \ldots g_{\sigma(k)}).
\end{equation}
The homomorphism induced by the standard inclusion 
${\rm gl}_{n-1}\rightarrow {\rm gl}_n$ sends
$\Phi^{n}$ into $\Phi^{n-1}$. The cochain $\Phi^n$ is a cocycle
\cite{Fuks}. Denote the cohomology class of the cocycle $\Phi^n$
by $\varphi^n$.
The tensor product in connection with algebra 
${\rm gl}_n (A)$ becomes  
$M_n({\mathbb C})\otimes A$. There are chain maps 
\begin{eqnarray}
\varphi^n :\, 
C^\bullet(A) & \longrightarrow &  
C^\bullet({\rm gl}_n (A),\, 
{\rm gl}_n(A)^*),
\\
\varphi^n(\tau)
(M_1\otimes a_1,\ldots, M_k\otimes a_k)
(M_0\otimes a_0)
& = & \sum_{\sigma\in S_k}
\mathrm{sign}(\sigma)
\tau(a_0\otimes a_{\sigma(1)}\dots\otimes a_{\sigma(k)})
\nonumber \\
&{}&
\times
\mathrm{Tr}(M_0 M_{\sigma(1)}\cdots M_{\sigma(k)}),
\label{maps}
\end{eqnarray}
where $\tau$ is the Hochschild cochain which is a ${\mathfrak s}{\mathfrak p}_{2n}({\mathbb C})$-invariant cocycle in the normalized Hochschild complex 
$
\overline{C}^{2n}({A}_{2n}^{\rm pol}) =
{A}_{2n}^{\rm pol}\otimes 
({A}_{2n}^{\rm pol}/{\mathbb C}\cdot{\id})^{\otimes 2n}
$
\cite{Feigin}.
These maps are compatible with the inclusion 
$\iota_{n'n}: {\rm gl}_n \rightarrow {\rm gl}_{n'}$,
$n < n'$ obtained by embedding an $n \times n$ matrix in
the first rows and columns of an $n'\times n'$ matrix
and completing with zeros. Indeed $\iota_{n' n}$ induces
a restriction map $\iota_{n' n}^*$ on complexes and
one has
$
\varphi^{n}=\iota_{n' n}^*\circ \varphi^{n'}.
$
Recall that $SV =\oplus S^jV$ denote the symmetric algebra
of a $A$-module $V$, and by composing $\varphi^n$ with the dual of the homomorphism of 
${\rm gl}_n(A)$-modules (compare with relations in Theorem \ref{TM})
one gets
\begin{equation}
S{\rm gl}_n (A)
\longrightarrow {\rm gl}_n (A),\,\,\,\,\,\,\,\,\,
z_1\cdots z_k \longmapsto \frac{1}{k!}
\sum_{\sigma\in S_k}z_{\sigma(1)}\cdots z_{\sigma(k)}\,,
\end{equation}
where the product on the left side is the product in the symmetric algebra of the vector space $S{\rm gl}_n(A)$, while the product on the right side is the associative product of $M_n({\mathbb C})\otimes {A}$.
Extensions of $\varphi^n$ to a chain map for all 
$j\geq 1$ are 
\begin{equation}
\varphi^n_j\,:\, C^\bullet(A) 
\longrightarrow  C^\bullet({\rm gl}_n(A),\, 
S^j{\rm gl}_n(A)^*).
\label{map}
\end{equation}

\subsection{The Hochschild cohomology}

Recall the definition of the Hochschild cohomology. Let $\cA$ be an associative algebra over ${\mathbb C}$. The Hochschild cochain complex with coefficients in $\cA$ is the sequence of vector spaces
$
C^n(\cA)= {\rm Hom}_{\mathbb C}(\cA^{\otimes n}, \cA),\, n=0,1,\ldots
$,
equipped with an operator $d_{\rm Hoch}: C^n(\cA)\rightarrow C^{n+1}(\cA)$,
\begin{eqnarray}
(d_{\rm Hoch} f)(a_1,\ldots,a_{n+1}) & = &
a_1 f(a_2,\ldots,a_n)
\nonumber \\
& + & \sum_{j=1}^n (-1)^j f(a_1,\ldots,a_{j-1},
a_j a_{j+1},a_{j+2},\ldots,a_n)
\nonumber \\
& + &  (-1)^{n+1} f(a_1,\ldots,a_n) a_{n+1}\,, 
\end{eqnarray}
where $d^2_{\rm Hoch} = 0$.
The cohomology of $d_{\rm Hoch}$ in degree $n$ will be denoted ${HH}^n(\cA)\equiv {HH}^n(\cA, \cA)$,
\begin{equation}
{HH}^n(\cA):=\frac{{\rm Ker}(d_{\rm Hoch}: C^n(\cA)\longrightarrow C^{n+1}(\cA))}
{{\rm Im} (d_{\rm Hoch}: \, C^{n-1}(\cA)\longrightarrow C^n(\cA))}
\end{equation}
and is called {\it the Hochschild cohomology of $\cA$ with coefficients in $\cA$}. Suppose that $\cA$ is a ${\mathbb Z}$-graded 
algebra and $\cA_p$ is a degree-$p$
component of $\cA$, such that $\cA_p\cdot \cA_q \subset \cA_{p+q}$.
We say that element $f$ of $C^n(\cA)$ has an internal degree $p$ if
$f(a_1,\ldots , a_n)\in \cA_{p+k_1+\cdots+k_n},
$
$a_i\in \cA_{k_i}$. The vector space $C^n(\cA)$ is graded by the internal degree, and the total degree of an element has the form
$C^\star(\cA) = \oplus_n C^n(\cA)$.
The Hochschild complex is graded by the total degree, and the
Hochschild differential can be expressed in the form
\begin{eqnarray}
(d_{\rm Hoch} f)(a_1,\ldots,a_{n+1}) & = &
(-1)^{a\cdot f}\ a_1 f(a_2,\ldots,a_n)
\nonumber \\
& + & \sum_{j=1}^n (-1)^j f(a_1,\ldots,a_{j-1},
a_j a_{j+1},a_{j+2},\ldots,a_n)
\nonumber \\
& + &  (-1)^{n+1} f(a_1,\ldots,a_n) a_{n+1}.
\end{eqnarray}

Now let ${\mathfrak A}=(\cA,Q)$ be a DG-algebra. The degree-1 derivation $Q$ as a map $Q:\, \cA_p\rightarrow \cA_{p+1}$
satisfies $Q^2=0$, and is given by
\begin{eqnarray}
(Qf)(a_1,\ldots,a_n) & = & Q(f(a_1,\ldots,a_n))
\nonumber \\
& - &
\sum_{j=1}^n (-1)^{v_1+\ldots+v_{j-1}+f+n-1}
f(a_1,\ldots,a_{j-1},Qa_i,a_{j+1},\ldots,a_n).
\label{Qf}
\end{eqnarray}
Each two-cocycle $({Q} f)(a_1,a_2)$ in (\ref{Qf}) defines an infinitesimal deformation of the associative product on $A$. 
Indeed, define a new product by
$
\alpha\circ \beta = \alpha \beta +t ({Q}f)(\alpha, \beta),\, t\in {\mathbb C};
$
then it will be associative of linear order in $t$ iff 
$({Q}f)=0$. Trivial infinitesimal deformations which lead to an isomorphic algebra are classified by two-coboundaries (i.e. two-cocycles of the form $({Q}f)\tau$ for some one-cochain $\tau(\alpha)$). 

Thus ${HH}^2(\cA)$ classifies nontrivial deformations of the associative algebra structure on $\cA$. A similar interpretation can be given to the Hochschild cohomology ${HH}^{\bullet}(\cA)$: it classifies infinitesimal deformations of $\cA$ in the class of $\cA_\infty$ algebras (associative algebras being a very special case of $\cA_\infty$ algebras).

The maps (\ref{map}) induce maps $\varphi^n_{k, j}$ on cohomology
\footnote{For a holomorphic function $W$ on $X$ one can define a deformed Lagrangian submanifold ${Y}$ by $p_i = \partial W/\partial q^i$. In symplectic geometry the function $W$ is known as the generating function of the Lagrangian ${Y}$.
Let us consider the polynomial Weyl algebra ${\cA}^{\rm pol}_{2n}$ over the
ring ${\mathbb C}[\varepsilon,\varepsilon^{-1}]$ following the lines of \cite{Feigin}. It is the space of polynomials 
$A[[p_1,\ldots,p_n, q_1,\ldots,q_n]]$ with the Moyal product 
$g_1\star g_2 = m(\exp(\varepsilon\alpha)(g_1\otimes g_2))$ ($m(g_1\otimes g_2)=g_1g_2$ being the standard commutative product 
on polynomial) which associated with the bivector
$
\alpha=(1/2)\sum_{i=1}^n\left(
\partial/\partial {p_i}\otimes\partial/\partial {q_i}
-
\partial/\partial {q_i}\otimes\partial/\partial {p_i}
\right)\in {\rm End}_{\mathbb C}({\cA}^{\rm pol}_{2n} \otimes {\cA}^{\rm pol}_{2n}).
$
There are defining relations $p_i\star q_j- q_j\star p_i=\varepsilon \delta_{ij}$ ($\hbar= i\varepsilon$ in
the notation of physics), and the following isomorphism:
\begin{eqnarray}
\!\!\!\!\!\!\!\!\!\!
&&
\left\{\!\!
\begin{array}{ll} 
{\rm Polynomial \,\,\,Weyl\,\,\, algebra}
\\
{\rm generated\,\,\, by} \,\,\,p_i, q_i
\end{array} 
\!\!\right\}
\stackrel{{\rm Isomorphism}}{\longleftarrow\longrightarrow}
\left\{\!\!
\begin{array}{ll} 
{\rm Algebra \,\,\, of\,\,\, differential\,\,\, operators\,\,\,in}
\,\,\, n\,\,\, {\rm variables}
\\
{\rm with\,\,\, coefficients\,\,\, in}\,\,\, 
{A}[[x_1,\ldots,x_{2n}]]
\end{array}
\!\!\right\} 
\nonumber
\end{eqnarray}  
If $\cA$ is the algebra of differential operators {\rm (}or the polynomial Weyl algebra{\rm )}, then
$\varphi^n_{k, j}$ is an isomorphism for all $0\leq k\leq n$ and $j\geq 1$.  In particular,
$
H^{k}({\rm g}{\rm l}_n(\cA_{2\ell}^{\rm pol}),\, 
S^j{\rm g}{\rm l}_n(\cA_{2\ell}^{\rm pol})^*)
= {\mathbb C}[\varepsilon,\varepsilon^{-1}]
$
for $k=2\ell$ (for $k<2\ell$ these cohomology groups are trivial).
}
:
\begin{equation}
\xymatrix@=15mm{
C^\bullet(\cA) 
\ar[r]^{\!\!\!\!\!\!\!\!\!\!\!\!\!\!\!\!\!
\!\!\!\!\!\!\!\!\!\varphi^n_j}
\ar@{-}[d]
&C^\bullet({\rm gl}_n(\cA),\, 
S^j {\rm gl}_n(\cA)^*)
\ar@{-}[d]
\\
{HH}^k(\cA)
\ar@{->}[r]^{\!\!\!\!\!\!\!\!\!\!\!\!
\!\!\!\!\!\!\!\!\!\!\!\!\!\varphi^n_{k, j\, (j\geq 1)}}
&H^k({\rm gl}_n(\cA),\, 
S^j {\rm gl}_n(\cA)^*)}
\label{filter11}
\end{equation}
\begin{remark}
From {\rm (\ref{commutator})} it follows that 
$
C^\bullet(W_n, {\rm g}{\rm l}_n; A) \hookrightarrow C^\bullet(W_n\ltimes {\overline G}\otimes P_n, {\rm g}{\rm l}_n; A).
$
Also there is an isomorphism of cohomology rings of reduced relative Weyl algebra and relative cochain complex of 
infinite dimensional Lie algebra $W_n\ltimes {\overline G}\otimes {P}_n$ over module of any Lie 
subalgebra ${\mg}\subset ({\rm g}{\rm l}_n\oplus{\overline G})$
{\rm (}see Eq. {\rm (\ref{D1})}{\rm )}:
$[\overline{\alpha}]: \, H^\bullet\widetilde{W}^\bullet
({\rm g}{\rm l}_n\oplus{\overline G}, {\mg})
\stackrel{\simeq}{\longrightarrow} H^\bullet 
(W_n\ltimes {\overline G}\otimes {P}_n, {\mg}; A)\,.
$
Note that the isomorphism $[\overline{\alpha}]$ does not depend on the choice of connection form $\alpha$. 
Thus we assume that the modified action of a topological model is  BRST-invariant and the deformed BRST charge is
$
Q = \overline{\partial} + \partial_{\rm deform},
$
where the operator $\partial_{\rm deform}$ is a linear vector field. Its Lie algebra is ${\rm gl}_n$ and the elements of 
this algebra in coordinates $z_1, \ldots , z_n$ is given in Eq. {\rm (\ref{LAlgebra})}. 
The operator $Q = \overline{\partial} + \partial_{\rm deform}$ satisfies the condition $Q^2 = 0$ 
and on the vector space $C^\bullet(\cA)$ there are two commuting differentials: $Q$ and $d_{\rm Hoch}$. 
The Hochschild cohomology of $A$ is defined to be the cohomology of 
$(-1)^n (\overline{\partial} + \partial_{\rm deform})+d_{\rm Hoch}$.
\end{remark}
\begin{conjecture} 
It has been shown that closed topological string states are related to infinitesimal deformations of the open-string theory. 
The closed string correlators perhaps can be constructed from the open ones using topological string theories as a model. 
The conjecture is 
{\rm \cite{Kapustin1}}: the space of physical closed-string states is isomorphic to the Hochschild cohomology of $(\cA,Q)$. This conjecture has been partially verified 
by means of computation of the Hochschild cohomology in the category of D-branes.

One can generalize this statement for the case of perturbative deformations {\rm (}see also {\rm \cite{BytsenkoProceedings}}{\rm )}. 
If there exists in the theory a single D-brane then all the information associated with deformations is encoded in an associative 
algebra $\cA$ equipped with a differential ${Q} = \overline{\partial} + \partial_{\rm deform}$. Equivalence classes of deformations 
of these data are described by a Hochschild cohomology of $(\cA, Q)$, an important geometric invariant of the 
{\rm (}anti{\rm )}holomorphic structure on $X$.
\end{conjecture}

\section{Deformation pairing}

{\bf The harmonic structure.} 
Suppose that $X$ is a smooth manifold of dimension $n$ over $\mathbb C$. The vector space structure of the harmonic structure 
$(HT^i(X);\,\, H\Omega_i(X))$ of $X$ is defined as
\begin{equation}
HT^i(X) = \bigoplus_{p+q=i} H^p(X, \Lambda^q TX)\,,
\,\,\,\,
H\Omega_i(X)  = \bigoplus_{q-p=i} H^p(X, \Omega^qX).
\label{harmonic}
\end{equation}
These vector spaces carry the same structures as 
$(HH^i(X);\,\, HH_i(X))$, namely:
$HT^i(X)$ is a ring, with multiplication induced by the exterior product on polyvector fields; 
$H\Omega_i(X)$ is a module over
$HT^i(X)$, via contraction of polyvector fields with forms.
It follows that (compare with results of Section 2):  
\begin{itemize}
\item{}
$HH^2(X)$ contains $H^1(X,TX)$, the space of infinitesimal complex structure deformations. 
\item{}
$H^0(X,\Lambda^2 TX)$ is a global bivector, giving rise to a noncommutative deformation.
\footnote{
In the Seiberg-Witten limit, we take the matrix inverse of the B-field to obtain the noncommutative deformation $\theta$, \cite{Seiberg}, which is this bivector. This is equivalent to inverting a symplectic form to obtain a Poisson structure.
} 
\item{}
The group $H^2(X,\cO)$ is a Gerbry deformation (see also Eq. (\ref{deform11})). 
\end{itemize}
For a smooth proper variety $X$ $({\rm dim}_{\mathbb C}X = n)$ we will use the following notation: 

- The diagonal embedding
$
\triangle: \, X\hookrightarrow X\times X= X^2,
$
\,
$K_X$ is the canonical bundle of $X$.

- D$^b(\cC)$ is equivalent to the full subcategory of 
D$(\cC)$ consisting of objects $X$ such that $H^n(X)=0$
for $|n|\gg 0$.

- $S_X = K_X$ is the dualizing object of 
D$_{\rm coh}^b(X)$,\, also thought to be as the Serre functor
$S_X\otimes (-) = K_X[n]\otimes (-)$;\,
$S_{X^2}= K_{X^2}({\rm dim}X^2 = 2n)$ in D$_{\rm coh}^b(X^2)$.

- $\cO_\triangle = \triangle_*\cO_X$ (the structure sheaf of the diagonal in $X\times X$);\,
$S_\triangle=\triangle_*S_X$;\,
$S\triangle^{-1}=\triangle_*S_X^{-1}$; \, the left adjoint of $\triangle_*$ is $\triangle_!:$ D$_{\rm coh}^b(X)\rightarrow$ 
D$_{\rm coh}^b(X^2)$;\,
$\triangle_! = S_{X^2}^{-1}\triangle_*S_X$ (also note that
$\triangle_!\cO_X \cong S_\triangle^{-1}$).
The identity functor from D$(X)$ to itself is given by the kernel $\Delta_*\cO_X = \cO_\Delta$ which is a coherent sheaf. 
We will refer to a sheaf of $\cO_X$-modules as an $\cO_X$-module. One can take a some of copies $\cO_X^{\oplus n} = (\underbrace{\cO_X \oplus \cO_X \oplus \ldots \oplus \cO_X}_n$
to give another $\cO_X$-module (the free $\cO_X$-module of rank $n$). 

The identification $\bigoplus_{p+q = n}H^p(X, \Lambda^qTX)$ with the group 
$\Ext_{X^2}^n(\cO_\Delta, \cO_\Delta)$ 
is given by the HKR isomorphism (see Appendix C) between these two groups,
\begin{eqnarray}
HH^i(X) & \stackrel{{\rm def}}{=}&  HH^i(D(X)) = {\rm Hom}_{D(X^2)}(\cO_\Delta,\cO_\Delta[i]) = {\rm Ext}^i_{X^2}(\cO_\Delta,\cO_\Delta) 
\nonumber \\
& = & {\rm Hom}_{D(X)}(\mathbf{L}\Delta^*\cO_\Delta,\cO_X[i]) = {\rm Ext}_X^i(\mathbf{L}\Delta^*\cO_\Delta,\cO_X)\ .
\label{isom1}
\end{eqnarray}
The reader can recognize this result as the space of closed string states in the B-model.

{\bf Algebra deformations.}
The HKR theorem states that for a commutative algebra $\cA$, 
$
HH^i(\cA) \cong  \Lambda^i{\rm Der}(\cA),
$ 
where Der$(\cA)$ is the space of derivations of $\cA$. There is a map from $\cA$ to ${\rm Der}(\cA)$ which gives rise to an 
exact sequence of algebra homomorphisms
\begin{equation}
0 \longrightarrow {A}\longrightarrow \cA \stackrel{\mathfrak m}{\longrightarrow}{\rm Der} (\cA)\longrightarrow 0
\end{equation}
where $\mathfrak m$ is the map $({v} + g)\mapsto [{v} + g,\, 
-]$,\,  ${v}\in {\rm gl}_n,\, g\in {\overline G}$.
The equivalent statement is:
$
HH^i({\rm Spec}(\cA)) \cong \, H^0({\rm Spec}(\cA),\Lambda^i T{\rm Spec}(\cA)),
$
where ${\rm Spec} (\cA)$ is the set of all proper prime ideals of $\cA$. 
\footnote{
For example, if $\chi$ is a character of $\cA$ (i.e. a non-zero homomorphism $\cA\rightarrow {\mathbb C}$) then ${\rm ker}\,\chi = 
\{ a\in \cA: \chi(a) = 0\}$ is an ideal of $\cA$. For ${\rm codim}\, \chi =1$ the ideal is maximal.
}
Since any variety can be covered by affine patches, one can think of the HKR theorem as a globalization of this result.

Let us consider a geometrical interpretation of the Hochschild cohomology. One can regard an associative algebra $\cA$ as the 
algebra of functions on an affine scheme $X={\rm Spec}(\cA)$.
Then consider $\cA\otimes \cA$, its spectrum ${\rm Spec}(\cA\otimes \cA)=X^2$, and the diagonal $\triangle\subset X^2$.
One can analyze open-string spectrum of $\triangle$ (i.e. the endomorphism algebra); it turns out that the resulting algebra of 
physical open-string states is precisely the Hochschild cohomology of $\cA$. Indeed, the Hochschild cohomology of $\cA$
is 
$
HH^i(\cA)= {\rm Ext}_{\cA\otimes \cA}^i(\cA, \cA);
$
it is the endomorphism algebra of $\cA$ regarded as an object of the derived category of modules over $\cA\otimes \cA$.
\begin{remark}
If $\cA$ is noncommutative, then $\cA$ is not a module
over $\cA\otimes \cA$, but it is a module over $\cA\otimes \cA^{\rm op}$, where $\cA^{\rm op}$ is the opposite
algebra of $\cA$. Thus we will have the more general case:
$
HH^i(\cA)= {\rm Ext}_{\cA\otimes \cA^{\rm op}}(\cA, \cA)
$
for which we need to compute the endomorphisms of $\triangle$ in D$^b(X^2)$. That is, one has to take a projective
resolution of $\cA$ regarded as a module over $\cA\otimes \cA^{\rm op}$, apply to it the operation
${\rm Hom}_{\cA\otimes \cA^{\rm op}}(-, \cA)$, and evaluate the cohomology of the resulting complex of vector spaces.
The main point is that for {\it any algebra} $\cA$ with a unit there is a canonical resolution of $\cA$
by free $\cA\otimes \cA^{\rm op}$ modules:
$
\cdots \rightarrow \cA^{\otimes 4}\rightarrow \cA^{\otimes 3}
\rightarrow \cA^{\otimes 2}.
$
Each term in this complex is a bimodule over $\cA$, which is the same as a module over $\cA\otimes \cA^{\rm op}$.
Then if we use this resolution to compute ${\rm Ext}^i(\cA, \cA)$, we get the Hochschild complex.
\end{remark}
For an affine space $X$ we have the following diagram 
\begin{equation}
\xymatrix @C=10mm{
&HH^i({\rm Spec (\cA)})
\ar[dl]_{\varphi^n_{k, j}}\ar[dr]^{\cong}&\\
H^i({\rm gl}_n(\cA),\, S^j{\rm gl}_n(\cA)^*)
\ar@{--}[r]&&
H^0({\rm Spec} (\cA),\, \Lambda^iT{\rm Spec}(\cA))
\ar@{--}[l]\\
}
\label{D2}
\end{equation}
\\
In the case of the B-model a bulk-boundary OPE would be a pairing
\begin{equation}
H^0(X,\, \Lambda^iTX) \times {\rm Ext}_X^n(\cE, \cF)\rightarrow
H^i({\rm gl}_n(\cA),\, S^j{\rm gl}_n(\cA)^*) \times {\rm Ext}_X^n(\cE, \cF) \rightarrow {\rm Ext}_X^{n+i}(\cE, \cF)
\end{equation}
\begin{conjecture} It is natural to conjecture that such a mathematical pairing realizes some bulk-boundary OPE above under 
deformation of the BRST operator. Perhaps any mathematical {\rm (}or geometrical\,{\rm )} deformations of a sheaf match 
physical deformations of the corresponding branes in the sigma model.
\end{conjecture}
In the special case that the sheaves are bundles on $X$,
so that the Ext groups reduce to sheaf cohomology on $X$, represented by differential forms, the mathematical pairing reduces 
to a wedge product of differential forms, similar to the Yoneda pairing in such circumstances, and the bulk-bulk OPE in all circumstances. This bulk-boundary pairing is defined as follows:

- First, we must define a map 
$
HH^i(X)\rightarrow H^i({\rm gl}_n(\cA),\, S^j{\rm gl}_n(\cA)^*) \rightarrow \mbox{Ext}_X^{i}( {\cE}, {\cE})
$
that maps bulk states (in the presence of deformations) to states defined on the boundary.

- Then, we use the Yoneda pairing 
$
\mbox{Ext}_X^{i}\left( {\cE}, {\cE} \right) \times
\mbox{Ext}^n_X\left( {\cE}, {\cF} \right) \rightarrow
\mbox{Ext}^{n+i}_X\left( {\cE}, {\cF} \right).
$

For the first map we need to identify $HH^i(X)$ (and therefore 
$H^i({\rm gl}_n(\cA); S^j{\rm gl}_n(\cA)^*)$) with the group 
$\Ext_{X^2}^i(\cO_\Delta, \cO_\Delta)$, see Eq. (\ref{isom1}).
Then we need define a pairing on
$H\Omega_i(X)$ which is a modification of the usual pairing of forms given by cup product and integration on $X$.  
Given the morphism above, we can define
the desired bulk-boundary map. 
Note that a Fourier-Mukai transform with kernel
${\cO}_{\Delta}$ maps ${\cE}$ to ${\cE}$, and with kernel
${\cO}_{\Delta}[n]$ maps ${\cE}$ to ${\cE}[n]$.
A bulk state identified via the morphism above as an element of 
$
\mbox{Ext}^n_{X^2} \left( {\cO}_{\Delta}, 
{\cO}_{\Delta} \right)
= \mbox{Hom}_{X^2}\left( {\cO}_{\Delta},
{\cO}_{\Delta}[n] \right)
$
is a map ${\mu}: {\cO}_{\Delta} \rightarrow {\cO}_{\Delta}[n]$.
Finally a map between the kernels of two Fourier-Mukai transforms
defines a map ${\cE} \rightarrow {\cE}[i]$ between the image
of a given object (${\cE}$), and the bulk state ${\mu}$ defines an element of
$
\mbox{Hom}_X\left( {\cE}, {\cE}[n] \right) = 
\mbox{Ext}^n_X\left( {\cE}, {\cE} \right).
$
Given an element of $H^i({\rm gl}_n(\cA),\, S^j{\rm gl}_n(\cA)^*)$ one can define an element of $\mbox{Ext}^{p+i}_X\left( {\cE},{\cE} \right)$.

\subsection*{Acknowledgments}

A. A. Bytsenko would like to acknowledge the Conselho Nacional
de Desenvolvimento Cient\'ifico e Tecnol\'ogico (CNPq, Brazil) and Funda\c cao Araucaria 
(Parana, Brazil) for financial support. The support of the Academy of Finland under the Project No. 136539 and 140886 is gratefully acknowledged. We thank Prof. Izumi Ojima for several useful discussions.

\section{Appendix A: Expository remarks on Lie algebra cohomology }

Fix a Lie algebra ${\mathfrak g}$ and a vector space $A$ over a field $F$.
In general both ${\frak g}$ and $A$ could be {\it infinite dimensional}.
Given representation $\pi: \, {\mathfrak g} \rightarrow {\mathfrak g}{\mathfrak l}(A)$
of ${\frak g}$ on $A$ we recall the definition of the $n^{th}$-dimensional cohomology $H^n({\frak g}; A)$ of ${\frak g}$
with coefficients in $A$ (for $n\geq 0$ an integer). The space of $n$-dimensional cochains $\Lambda^n({\frak g}; A)$ is define as follows.
$\Lambda^0({\frak g}; A) \stackrel{{\rm def}}{=} A$, and for $n\geq 1$,
$\Lambda^n({\frak g}; A)$ is the space of $n$-linear maps $f: \, 
{\frak g}\times \cdots \times {\frak g} \rightarrow A$ that are {\it alternating}; i.e. $f(x_1, \cdots, x_i, \cdots , x_j, \cdots , x_n) = 0$
for $i\neq j, \, x_i = x_j, x_i \in {\frak g}$.

Assuming that the characteristic of $F$ is not 2 (say $F = {\mathbb R}$ or $\mathbb C$) these alternating maps are the same as skew-symmetric maps. We have a linear map $\delta_n : \, \Lambda^n({\frak g}; A) \rightarrow \Lambda^{n+1}({\frak g}; A)$ 
(a coboundary operator) defined by 
\begin{eqnarray}
\!\!\!\!\!\!\!\!\!\!\!\!\!
(\delta_n, f)(x_1, x_2, \ldots ,x_n, \ldots , x_n, x_{n+1})\!
& \stackrel{{\rm def}}{=} & 
\!\!\!\sum_{i=1}^{n+1}(-1)^{i+1}\pi(x_i)f(x_1, \ldots,  \hat{x}_i, \ldots , x_{i+1})
\nonumber \\
\!
& + &
\!\!\!\sum_{i< j}(-1)^{i+j} f([x_i, x_j], x_1, \ldots , \hat{x}_i, \ldots,
\hat{x}_j, \ldots , x_{n+1}),
\end{eqnarray}
for $n\geq 1$, $(\delta_0 v)(x)\stackrel{{\rm def}}{=} \pi(x)v$ for $v\in A = \Lambda ^0({\frak g}, A)$. For example 
\begin{equation}
\delta_1 : \, \Lambda^1({\frak g}; A) \stackrel{{\rm def}}{=} 
{\rm Hom}({\frak g}; A)\longrightarrow \Lambda^2({\frak g}; A)
\end{equation} 
is given by 
\begin{equation}
(\delta_1 f)(x_1, x_2)\stackrel{{\rm def}}{=}\pi(x_1)f(x_2) -
\pi(x_2)f(x_1) - f([x_1, x_2])\,,
\label{d1}
\end{equation} 
for $f : \, {\frak g}\rightarrow A$ in Hom$({\frak g}; A)$.
Similarly  \, $\delta_2 : \, \Lambda^2({\frak g}; A) \rightarrow 
\Lambda^3({\frak g}; A)$ is given by 
\begin{eqnarray}
(\delta_2 f)(x_1, x_2, x_3) 
& \stackrel{(\rm def)}{=} & 
\pi(x_1)f(x_2, x_3) - \pi(x_2)f(x_1, x_3) + \pi(x_3)f(x_1, x_2) - f([x_1, x_2], x_3) 
\nonumber \\
& - &
f([x_2, x_3], x_1) + f([x_1, x_3], x_2)\,.
\label{d2}
\end{eqnarray}
Then by direct calculation, $\delta_1\delta_0 = 0, \, \delta_2\delta_1 = 0$. In general one can prove that $\delta_{n+1}\delta_n :
\Lambda^n({\frak g}; A)\rightarrow \Lambda^{n+2}({\frak g}; A)$ is zero map. 
\begin{definition}
For $n\geq 1$, $B^n({\frak g}; A)\, \stackrel{\rm def}{=}$
{\rm Im} {\rm (}image{\rm )} of $\delta_{n-1}\subset$ {\rm Ker} 
{\rm (}kernel{\rm )} of $\delta_n\stackrel{{\rm def}}{=} Z^n({\frak g}; A)$ and we can define 
\begin{equation}
H^n({\frak g}; A)\, \stackrel{{\rm def}}{=} Z^n({\frak g}; A) /B^n({\frak g}; A)\,.
\end{equation}
\end{definition}
We set $H^0({\mathfrak g}; A) = {\rm Ker}\, \delta_0 = \{v\in A\, \mid \pi(x)v = 0,\, \forall x \in {\mathfrak g}\}\, \stackrel{{\rm def}}{=} A^{\mathfrak g}$  (= space of {\it invariants}). Elements of $B^n({\mathfrak g}; A), Z^n({\mathfrak g}; A)$ are called $n$-{\it coboundaries}, $n$-{\it cocycles}, respectively (with coefficients in $A$). In particular for $A = {\mathfrak g}$, with $\pi = {\rm ad}_{\mathfrak g}$ = the {\it adjoint representation} of $\mathfrak g$ on $\mathfrak g$
(i.e. $\pi (x) = {\rm ad}_x : \, {\mathfrak g} \rightarrow 
{\mathfrak g}$, with ${\rm ad}_x (y) \stackrel{\rm def}{=} [x, y]$) the group $H^2({\mathfrak g}, {\mathfrak g})$ classifies the {\it infinitesimal deformations} of ${\mathfrak g}$ (up to equivalence), by a result due to M. Gerstenhaber \cite{Gerstenhaber64,Gerstenhaber66,Gerstenhaber68}.

\subsection{Examples}
{\bf The Witt algebra.} 
Let ${\mathfrak g}$ = the complexified Lie algebra of polynomial vector fields on the circle $S^1$: ${\mathfrak g} = W$ = the Witt algebra spanned (as a vector space) by $\{L_n\}_{n\in {\mathbb Z}}$ with
$[L_n, L_m] = (n-m) L_{n+m}$. Let $\pi$ be the {\it trivial} representation of ${\mathfrak g}$ on $A = {\mathbb C} : \, \pi(x)z
\stackrel{\rm def}{=} 0, \,\, \forall (x, z)\in {\mathfrak g} \times {\mathbb C}$. Then it is due to Gelfand-Fuks \cite{Gelfand68} that $H^2({\mathfrak g}; {\mathbb C}) = {\mathbb C}$. In fact one can construct $\omega \in Z^2({\mathfrak g}, {\mathbb C})/ 
B^2({\mathfrak g}, {\mathbb C})$ by $\omega(L_n, L_m)
\stackrel{\rm def}{=} \delta_{n+m,0}\,n(n²-1)/12$, where the factor $1/12$ (the value of the Riemann zeta function at -1) is only a convenient normalization.
A deeper (deformation) result is due to A. Fialowski \cite{Fialowski}, which shows that ${\mathfrak g}$ is {\it rigid}, infinitesimally and formally: $H^2({\mathfrak g}, {\mathfrak g}) = 0$.

{\bf The Virasoro algebra.} Using the cocycle $\omega$ above, one constructs the {\it Virasoro algebra} ${\mathfrak g}$ = Vir as a {\it central extension} of $W$. That is, one constructs an exact sequence of Lie algebras
\begin{equation}
0\longrightarrow A\longrightarrow {\mathfrak g}\longrightarrow W\longrightarrow 0 
\end{equation}
with $A$ abelian $ \ni [A, {\mathfrak g}] = 0$. Namely take 
$A$ = 1-dimensional vector space ${\mathbb C}z$, and 
Vir$\stackrel{\rm def}{=} W\times {\mathbb C}z$, with bracket \begin{equation}
[(\omega_1, c_1z), \, (\omega_2, c_2z)] \stackrel{\rm def}{=}
([\omega_1, \omega_2], \, \omega(\omega_1, \omega_2)z)\,\,\,\,\,
{\rm for}\,\,\,\,\,
(\omega_1, \omega_2, c_1, c_2)\in W\times W\times {\mathbb C}\times {\mathbb C}\,.
\end{equation}
For $\widehat{L}_n\stackrel{\rm def}{=}(L_n, 0), \, \widehat{z} =
(0, z)\in {\rm Vir}$, one obtains the familiar rules:
\begin{equation}
[\widehat{L}_n, \, \widehat{L}_m] = (n-m)\widehat{L}_{n+m} + \frac{n(n^2-1)}{12}\delta_{n+m, 0} \widehat{z}, \,\,\,\,\,\,\,\,
[\widehat{L}_n, \, \widehat{z}] = 0,\,\,\,\,\,\,\,\,
[\widehat{z}, \, \widehat{z}] =0\,.
 \end{equation}
The deformation result for $W$ also holds for Vir \cite{Fialowski03}: Vir is infinitesimally and formally rigid: $H^2({\rm Vir}, {\rm Vir}) =0$. 
Note that by formula (\ref{d2}),
\begin{equation}
\omega(x_3, [x_1, x_2]) + \omega(x_1, [x_2, x_3]) +
\omega(x_2, [x_3, x_1]) = 0\,. 
\end{equation}

{\bf Complex semisimple Lie algebra.}  Let $\mathfrak g$ be a complex semisimple Lie algebra which is finite dimensional. Let $\mathfrak h$
be a Cartan subalgebra of $\mathfrak g$, ${\mathfrak h}^* = {\rm Hom}
({\mathfrak h}, {\mathbb C})$, \, $W$ the Weyl group of 
$(\mathfrak g, \mathfrak h)$, and for a choice of positive root system $\triangle^+$ of $(\mathfrak g, \mathfrak h)$ let 
$\delta = (1/2)\sum_{\alpha \in \triangle^+}\alpha$. For $\lambda\in {\mathfrak h}^*$, let $M(\lambda)\stackrel{\rm def}{=} 
U({\mathfrak g})\bigotimes_{U({\mathfrak b})}{\mathbb C}_\lambda$ be the corresponding {\it Verma module} of $\mathfrak g$, where 
${\mathfrak b} \supset {\mathfrak h}$ is a Borel subalgebra of $\mathfrak g$ and where ${\mathbb C}_\lambda = {\mathbb C}$ as a $\mathfrak b$-module $\ni H\cdot z \stackrel{\rm def}{=} \lambda(H)z$, $[{\mathfrak b}, {\mathfrak b}]\cdot z \stackrel{\rm def}{=} 0$ for 
$(H, z)\in {\mathfrak h}\times {\mathbb C}$; $U({\mathfrak b})$ is the universal enveloping algebra of a Lie algebra $\mathfrak b$.
The representation  $\pi$ of $\mathfrak g$ on $M(\lambda)$ is given by 
$\pi(x)(u\otimes z) \stackrel{\rm def}{=} xu\otimes z$ \, for 
$(x, u, z) \in {\mathfrak g}\times U({\mathfrak g})\times {\mathbb C}$.
The cohomology groups $H^n({\mathfrak g}; M(\lambda))$ were computed by F. Williams \cite{Williams78}:
\begin{itemize}
\item{}
$H^n(\mg,\, M(\lambda)) = 0, \, \forall n\geq 0$\,, unless $\lambda = \omega\delta - \delta$ for some (necessarily unique) $\omega \in W$, in which case $H^n(\mg,\, M(\lambda)) = \Lambda^{{\rm dim}\,\mg -n - ({\rm length\,\, of}\,\,\omega)}\mh$\,.
\end{itemize}

\subsection{Lie algebra homology and spectral sequences}
\label{LAH}

It is of interest to compute also the Lie algebra {\it homology} $H_n(\mg,\, M(\lambda))$. More generally if $\mg$ is any Lie algebra and $\pi : \, \mg \rightarrow \mg\ml(A)$ is a representation of $\mg$ on $A$ then one has the following relation $H_n(\mg; A)^* = H_n(\mg; A^*)$ where the representation of $\mg$ on $A^* = {\rm Hom}(A, {\mathbb C})$ is the {\it contragradient} $\pi^*$ of $\pi$; that is $(\pi^*(x) f)(v)\stackrel{\rm def}{=} - f(\pi(x) v)$ for 
$(x, f, v) \in \mg\times A^*\times A $. Since $\mg$ here is arbitrary we can now take $\mg$ to be an (infinite dimensional)
Kac-Moody Lie algebra $\mg(A)$ associated to a symmetrizable generalized $\ell \times \ell$ Cartan matrix $A = [a_{ij}]$.
Thus the $a_{ij}$ are intergers with $a_{ij} = 2,\, a_{ij}\leq 0$ for $i \neq j$, and for some diaginal matrix $D$ with $\ell$ positive rational entries $DA$ is a symmetric matrix. Using ideas similar to those developed in \cite{Williams78}, C. Sen \cite{Sen84} and S. Kumar \cite{Kumar86}, independently, obtain the following result which is dual to that in complex semisimple Lie algebra example, for a Cartan subalgebra $\mh$ of $\mg$:
\begin{itemize}
\item{}
$H_n(\mg,\, M(\lambda)) = 0, \forall n\geq 0$\,, unless $\lambda = w\delta - \delta$ for some (necessarily unique) $w \in W$, in which case $H_n(\mg,\, M(\lambda)) = \Lambda^{n- ({\rm length\,\, of}\,\,w)}\mh$\,.
\end{itemize}
When $\mg(A) = \mg$ in the case of a finite dimensional complex semisimple Lie algebra (see above), this result reduces to that in \cite{Williams78}.
\begin{definition}
\label{Definition8.2}
A spectral sequence is a family of abelian groups $\{E_r^{p, q}\}$ 
and homomorphisms $\{d_r^{p, q}\}$ for $p, q, r \in {\mathbb Z}$, where 
$d_r^{p, q} : \, E_r^{p, q}\rightarrow E_r^{p+r, q-r+1}$ and where we require the following:
\begin{enumerate}
\item{}
The composition $E_r^{p-r, q+r-1}\rightarrow E_r^{p, q}\rightarrow E_r^{p+r, q-r+1}$ is the zero map. Hence 
${\rm Im}\, d_r^{p-r, q+r-1} \\
\stackrel{\rm def}{=} d_r^{p-r, q+r-1} E_r^{p-r, q+r-1} \subset
{\rm Ker} \,d_r^{p, q}$.
\item{}
${\rm Ker}\, d_r^{p-r, q+r-1}/{\rm Im}\, d_r^{p, q} \simeq E_{r+1}^{p, q}$
{\rm (}$\simeq$ means an isomorphism{\rm )}.
\end{enumerate}
\end{definition}

{\bf Spectral sequence of a filtered complex.} Let $C = \{C^n, d_n\}_{n \in {\mathbb Z}}$ be a cochain complex of abelian groups $C^n$,
\begin{equation}
C^{n-1}\stackrel{d_{n-1}}{\longrightarrow} C^n \stackrel{d_{n}}{\longrightarrow} C^{n+1}
\end{equation}
is the zero map and $H^n(C)\stackrel{\rm def}{=} 
{\rm Ker}\,d_n/d_{n-1}C^{n-1}$ is the $n^{th}$-dimensional cohomology of $C$. A {\it decreasing filtration} of $C^n$ is a family $\{F_pC^n\}_{p\in {\mathbb Z}}$ of subgroups $F_pC^n \subset C^n$ 
of $C^n$ $\ni$\, $F_{p+1}C^n\subset F_pC^n$\, $\forall\, p \in 
{\mathbb Z}$, and $d_nF_pC^n \subset F_pC^{n+1}$ \, $\forall\, p\in {\mathbb Z}$. We suppose that such a filtration exists $\forall\, n\in {\mathbb Z}$. 
\begin{definition}
We define abelian groups 
\begin{eqnarray}
Z_r^{p, q} & \stackrel{\rm def}{= } &
\{x\in F_pC^{p+q}\, \mid \, d_{p+q}x\in F_{p+r}C^{p+q+1}\}\,,
\\
B_r^{p, q} & \stackrel{\rm def}{= } &
d_{d+q-1} Z^{p-r, q+r-1}_r \equiv d_{p+q-1} F_{p-r}C^{p+q-1}\cap
F_pC^{p+q}\,.
\end{eqnarray}
Then $B_{r-1}^{p, q}$ and $Z_{r-1}^{p+1, q-1}$ are subgroups of $Z_r^{p, q}$
and we define a spectral sequence by first setting
\begin{equation}
E_r^{p, q} \stackrel{\rm def}{=} Z_r^{p,q}/(B_{r-1}^{p, q} +
Z_{r-1}^{p+1, q-1})\,. 
\end{equation}
\end{definition}
Note that $d_{p+q} : \, Z_r^{p, q}\rightarrow  Z_r^{p+r, q-r+1}$, 
$d_{p+q}$ maps $B_{r-1}^{p, q} \stackrel{\rm def}{=} d_{p+q-1} 
Z_{r-1} ^{p-r+1, q+r-2}$ to 0 (since $d_{n-1}d_n = 0$) and
\begin{equation}
B_{r-1}^{p+r, q-r+1}\stackrel{\rm def}{=} d_{p+q} Z_{r-1}^{p+1, q-1}
\Longrightarrow d_{p+q}:\, Z_{r-1}^{p+1, q-1}\longrightarrow B_{r-1}^{p+r, q-r+1}\,, 
\end{equation}
which shows that $d_{p+q}$ induces a quotient map 
$d_r^{p, q} : E_r^{p, q}\rightarrow E_r^{p+r, q-r+1}$  $\ni$
$d_r^{p, q} d_r^{p-r, q+r-1} \\ = 0$. To see that $\{E_r^{p, q}, d_r^{p, q}\}$ is a spectral sequence, one has to check that 
${\rm Ker}\,d_r^{p, q}/{\rm Im}\,d_r^{p-r, q+r-1} \\ \simeq E_{r+1}^{p, q}$\, $\forall\, p, q, r \in {\mathbb Z}$. 
In fact, if 
\begin{equation}
\pi : \, Z_r^{p, q}
\longrightarrow E_r^{p, q} \stackrel{\rm def}{=} Z_r^{p, q}/(B_{r-1}^{p, q} + Z_{r-1}^{p+1, q-1})
\end{equation} 
is the natural quotient map (where $z\rightarrow z + B_{r-1}^{p, q} + Z_{r-1}^{p+1, q-1}$ for $z\in Z_r^{p, q}$) then ${\rm Im}\,\pi (Z_{r+1}^{p, q})= {\rm Ker}\, d^{p, q}_r$ and an isomorphism $\Phi$ of $E_{r+1}^{p, q}$
onto ${\rm Ker}\, d_r^{p, q}/{\rm Im}\,d_r^{p-r, q+r-1}$ is given by 
\begin{equation}
\Phi(z+ B_r^{p, q} + Z_r^{p+1, q-1}) = \pi z + 
{\rm Im}\,d_r^{p-r, q+r-1}\,\,\,\,\, {\rm for}\,\,\,\,\,
z\in Z_{r+1}^{p, q}\,.
\end{equation}

{\bf Regularity.} To connect this spectral sequence with the computation of cohomology we assume that the filtration is {\it regular}: $\forall\, n \,\,\, \exists$ an integer $s(n) \ni F_pC^n =0$ for 
$p> s(n)$. In some cases in practice one can take $s(n) = n$. We also assume the {\it exhaustive} condition: $F^pC^n = C^n,\,\, \forall\,n$ when 
$p\leq 0$. One has
\begin{theorem} The following isomorphisms hold:

${(\mathbb A)}$  Let $r\geq 2 \, \ni$ for some $q_0,\,\, E_r^{p, q} = 0\,\,\,\,\, \forall p \,, \,\forall q \neq q_0$. Then $\forall p, \, H^p(C)\simeq E_r^{p-q_0, q_0}$.  

${(\mathbb B)}$  Let $r\geq 1 \, \ni$ for some $p_0,\,\, E_r^{p, q} = 0\,\,\,\,\, \forall q \,, \,\forall p \neq p_0$. Then $\forall\,p, \,H^p(C)\simeq E_r^{p_0, p-p_0}$.
\end{theorem}
Conditions ${\bf (\mathbb A), \,(\mathbb B)}$ mean that the spectral sequence {\it collapses} at $r$.

{\bf Spectral sequence of a double complex.} A {\it double complex} is a family of abelian groups $\{C^{p, q}\}_{p, q \in {\mathbb Z}}$ with homomorphisms
\footnote{
The $\partial$ are called {\it horizontal} operators and the $\overline{\partial}$ are called {\it vertical} operators.
}
:
\begin{eqnarray}
\!\!\!\!\!\!\!\!\!\!\!\!\!\!
\partial & : & \, C^{p, q}\longrightarrow C^{p+1, q}\,, \,\,\,\,\,
\,\,\,\,\,\,\, 
\overline{\partial} : \, C^{p, q}\longrightarrow C^{p, q+1}
\,\ni\, \partial\partial = \overline{\partial}\, \overline{\partial} = 0\,\,\,\,\, {\rm and} \,\,\,\,\, \partial\overline{\partial} + \overline{\partial}\partial = 0, 
\\
\!\!\!\!\!\!\!\!\!\!\!\!\!\!
\partial \partial & : & \, C^{p, q}\longrightarrow C^{p+2, q}\,, \,\,\,\,\,\,\,\,\,\,\,
\overline{\partial}\, \overline{\partial} : \, C^{p, q}\longrightarrow C^{p, q+2}\,,
\\
\!\!\!\!\!\!\!\!\!\!\!\!\!\!
\partial \overline{\partial} & : & \, C^{p, q}\longrightarrow C^{p+1, q+1}\,, \,\,\,\,\,
\overline{\partial} \partial  :  \, C^{p, q}\longrightarrow C^{p+1, q+1}\,.
\end{eqnarray}
One constructs a cochain complex $\{C^n, d_n\}_{n\in {\mathbb Z}}$ by
setting $C\stackrel{\rm def}{=} \sum_{p+q=n}C^{p, q}$ (direct sum),
$d_n= \partial + \overline{\partial}$ on $C^{p, q}\subset C^n \rightarrow C^{p+1, q} + C^{p, q+1} \subset C^{n+1}$. There are two decreasing filtrations of $C^n$, $\forall n \in {\mathbb Z}$ :
\begin{eqnarray}
F_p^1C^n  & \stackrel{\rm def}{=}&
\{ \sum_{r+s=n}x_{rs}\in C^n\,\mid \, x_{rs}\in C^{r,s},\, r\geq p\}\,, 
\\
F_p^2C^n  & \stackrel{\rm def}{=}&
\{ \sum_{r+s=n}x_{rs}\in C^n\,\mid \, x_{rs}\in C^{r,s},\, s\geq q\}\,. 
\end{eqnarray}
\begin{theorem} These filtrations therefore generate two spectral sequences ${}^{(j)}E_r^{p, q},\, j=1, 2$, whose initial terms are computed as follows:
\begin{eqnarray}
{}^{(1)}E_0^{p, q} & \simeq & C^{p, q}\,,\,\,\,\,\,
{}^{(2)}E_0^{p, q} \simeq C^{q, p}\,,
\\
{}^{(1)}E_1^{p, q} & \simeq & {\rm Ker}(C^{p, q}
\stackrel{\overline{\partial}}{\longrightarrow} C^{p, q+1})/\overline{\partial} C^{p, q-1}
\stackrel{\rm def}{=} H^q(C^{p, \bullet},\, \overline{\partial})\,,
\\
{}^{(2)}E_1^{p, q} & \simeq & {\rm Ker}(C^{q, p}
\stackrel{\partial}{\longrightarrow} C^{q+1, p})/\partial C^{q-1, p}
\stackrel{\rm def}{=} H^q(C^{\bullet, p},\, \partial)\,,
\end{eqnarray}
${}^{(1)}E_2^{p, q} \simeq {\rm Ker}\,\partial/{\rm Im}\,\partial$, where
\begin{equation}
\partial : \, {\rm Ker}(C^{p, q}
\stackrel{\overline{\partial}}{\longrightarrow} C^{p, q+1}) /\overline{\partial} C^{p, q-1}
\longrightarrow 
{\rm Ker}(C^{p+1, q}
\stackrel{\overline{\partial}}{\longrightarrow} C^{p+1, q+1}) /\overline{\partial} C^{p+1, q-1}
\end{equation}
is the quotient map induced by $\partial$; i.e. $\partial(x+\overline{\partial} C^{p, q-1})\stackrel{\rm def}{=} \partial x + \overline{\partial}C^{p+1, q-1}$ for $x\in C^{p,q}\, \ni\, \overline{\partial} x =0$, which is well define since $0 = \partial \overline{\partial} + \overline{\partial} \partial$ on $C^{p, q}$
$\Rightarrow \overline{\partial} \partial x = - \partial \overline{\partial} x = 0$. 
Similarly ${}^{(2)}E_2^{p, q} \simeq {\rm Ker}\,\overline{\partial}/{\rm Im}\,\overline{\partial}$, where
\begin{equation}
\overline{\partial} : \, {\rm Ker}(C^{q, p}
\stackrel{\partial}{\longrightarrow} C^{q+1, p}) /\partial C^{q-1, p}
\longrightarrow 
{\rm Ker}(C^{q, p+1}\stackrel{\partial}{\longrightarrow} C^{q+1, p+1}) /\partial C^{q-1, p+1}
\end{equation}
is the quotient map induced by $\overline{\partial}$ : $\overline{\partial}(x+ \partial C^{p, p})\stackrel{\rm def}{=} \overline{\partial} x + \partial C^{q-1, p+1}$
for $x\in C^{p,q} \ni \partial x =0$.
Suppose also that $C^{p, q} = 0$ for either $p< 0$ or $q< 0$. Then both filtrations are regular and exhaustive {\rm (}see previous section{\rm )}: in fact $F_p^{j=1, 2}C^n = 0$ for $p> n$ {\rm (}so $s(n) = n${\rm )}. Therefore the cohomology formulas $(\mathbb A),\, (\mathbb B)$ hold for collapse at $r\geq 0,\, r\geq 1$, respectively.
\end{theorem}

{\bf BRST cohomology.} Let $\mg$ be a finite dimensional Lie algebra over $\mathbb R$. Let $\Lambda^n \stackrel{\rm def}{=} \Lambda^n(\mg, {\mathbb R})$ be the space of $n$-cochains for the trivial  representation $\pi$ of $\mg$ on $A = {\mathbb R}$. Suppose we also have some other cochain complex $C = \{C^n, d_n\}_{n \in {\mathbb Z}}$. We assume in fact that each $C^n$ is a vector space over $\mathbb R$ and the $d_n : \, C^n\rightarrow C^{n+1}$ are linear maps over 
$\mathbb R$. We assume also that the $C^n$ carry a representation $\pi_n : \, \mg \rightarrow \mg\ml(C^n)$ of $\mg$ such that the diagram
\begin{equation}
\xymatrix@=15mm{
C^n 
\ar[r]^{d_n}
\ar[d]^{\pi_n(x)}
& C^{n+1}
\ar[d]^{\pi_{n+1}(x)}
\\
C^n
\ar@{->}[r]^{d_n}
& \, C^{n+1}}
\label{diagC}
\end{equation}
is commutative $\forall x\in \mg$; \, i.e. $d_n \pi_n (x) = \pi_{n+1}(x) d_n$. For example, $\mg$ could be the Lie algebra of a Lie group $G$ which has a Hamiltonian action on a symplectic manifold $(X, \omega)$, and we could take $C^n = \Lambda^n\mg\otimes C^{\infty}(X)$, as in the classical BRST setting.

One constructs a double complex by setting 
$C^{p, q}\stackrel{\rm def}{=} C^p\otimes \Lambda^q$ and by defining
\begin{eqnarray}
\partial & : & C^{p, q}\longrightarrow C^{p+1, q}\,,\,\,\,\,\,\,\,
\,\,\,\,
\overline{\partial} \,\,\,\,: \,\,\,\,\, C^{p, q}\longrightarrow C^{p, q+1}, 
\\
\partial & \stackrel{\rm def}{=} &
d_p\otimes \id_{\Lambda^\mg} \,,\,\,\,\,\,\,\,\,\,\,\,\,\,\,\,
\,\,\,\,\,\,\,\,\,\,\,\,
\overline{\partial} \,\,\, \stackrel{\rm def}{=} \,\,\, (-1)^p \id_{C^p} \otimes \delta_q\,,
\end{eqnarray}
where $\delta_q : \, \Lambda^q\rightarrow \Lambda^{q +1}$ is the coboundary operator and $\id_S$ is the identity operator on a space $S$. Already $C^{p, q} = 0$ for $q< 0$ and in the example just presented $C^{p, q} = 0$ for $p < 0$ which we need
\footnote{
Recall that as before this is needed for regularity and exhaustiveness of the filtrations.
}
- i.e. take $C^p = 0$ for $p < 0$. Of the two spectral sequences generated by this double complex we focus only on the second one with ${}^{(2)}E_1^{p, q} \simeq {\rm Ker}(C^{q, p}\stackrel{\partial}{\rightarrow} C^{q+1, p})/\partial C^{q-1, p}$.
The main point in the classical BRST setting is the acyclicity of $C$:
$H^n(C) = 0$ for $n \geq 1$, which allows for the collapse of the spectral sequence at $r = 1$. One computes that
\begin{equation}
{}^{(2)}E_1^{p, q}  \simeq  H^q(C)\otimes \Lambda^p = 0 \,\,\,\,
{\rm for} \,\,\,\, q \neq 0 \,\,  \Rightarrow
{}^{(2)}E_r^{p, q}  =  0 \,\,\,\, {\rm for} \,\,\,\, r>1 
\,\,\,
({\rm by}\,\,\, {\it 2} \,\,\, {\rm in}\,\,\,\, {\rm Definition}\,\,\, 8.2) \,. 
\end{equation}
In particular we have collapse at $r=2 : \,{}^{(2)}E_2^{p, q} =0$
for $q\neq 0$. Therefore for the total complex $C_{\rm total} \stackrel{\rm def}{=} \sum_{p+q = n}C^{p, q}$ with BRST operator $\partial + \overline{\partial}$, we deduce from $(\mathbb A)$ that 
\begin{equation}
H^p(C_{\rm total}) \simeq {}^{(2)}E_2^{p, q}\,, \,\,\,\,\, \forall p\,.
\end{equation}
Of interest is the zero-dimensional cohomology which one computes
(using previous results) as $H^0(C_{\rm total}) \simeq 
{}^{(2)}E_2^{0, 0} \simeq$ zero-dimensional $\overline{\partial}$-cohomology of ${}^{(2)}E_1^{0, q}$
(i.e. $H^0(H^0(C)\otimes \Lambda^\bullet; \overline{\partial})$)
$\simeq H^0(C)^q$, where that latter space of invariants make since as the commutative diagram (\ref{diagC}) leads to a representation $\pi_n$
of $\mg$ on the cohomology $H^n(C)$. The isomorphism $H^0(C_{\rm total})\simeq H^0(C)^q$ is classically the fact that invariant functions on a constraint manifold are described by BRST cohomology in degree zero \cite{Forger92}.

{\bf Relative cohomology.}
Suppose $\frak{h}$ is a subalgebra of the algebra $\frak{g}$. If $A$ is a $\frak{g}$-module then denote by $C^q(\frak{g}, \frak{h}; A)$ the subspace of the space $C^q(\frak{g},A)$,
consisting of cochains $c$, such that 
\begin{equation}
c(g_1, \dots , g_q)=0\,\,\,\, {\rm for}\,\,\,\, g_1\in \frak{h},
\,\,\,\, {\rm and}\,\,\,\, dc(g_1, \ldots , g_{q+1})=0\,\,\,\, 
{\rm for}\,\,\,\, g_1\in \frak{h}\,.
\end{equation} 
Equivalent definition: $C^q(\frak{g}, \frak{h}; A) = {\rm Hom}_{\frak{h}}(\Lambda (\frak{g}/\frak{h}), A)$. 
Elements of the space $C^q(\frak{g}, \frak{h}; A)$ are called {\it relative cochains}. Since 
$dC^q(\frak{g}, \frak{h}; A)\subset C^{q+1}(\frak{g}, \frak{h}; A)$ the relative cochains constitute a subcomplex of the 
complex $C^\bullet(\frak{g}, A)$. Denote this subcomplex by $C^\bullet(\frak{g}, \frak{h}; A)$; 
it cohomology $H^q(\frak{g}, \frak{h}; A)$ is called a (relative) cohomology of the algebra $\frak{g}$ modulo 
$\frak{h}$ with coefficients in $A$. For the base field $A$ we use the notation: $C^q(\frak{g}, \frak{h})$,
$H^q(\frak{g}, \frak{h})$. 
Let $\frak{h}$ be an ideal in $\frak{g}$. Then $\Lambda^q(\frak{g}/\frak{h})$ is trivial $\frak{h}$-module and 
\begin{eqnarray}
{\rm Hom}_{\frak{h}}(\Lambda^q(\frak{g}/\frak{h}), A) & = &
{\rm Hom}(\Lambda^q(\frak{g}/\frak{h}), {\rm Inv}_{\frak{h}}A)
= C^q(\frak{g}/\frak{h}; {\rm Inv}_{\frak{h}}A)\,,
\noindent \\
{\rm Inv}_{\frak h}A & = &  \{a\in A |\, ha=0 \,\,\,\, \forall \,h \in \frak{h}\}\,,
\end{eqnarray}
${\rm Inv}_{\frak{h}}A$ is the module over $\frak{g}/\frak{h}$ of $\frak{h}$-invariants. In this case the differentials in 
the complexes $C^\bullet(\frak{g}, \frak{h}; A)$ and $C^\bullet(\frak{g}/\frak{h}, {\rm Inv}_{\frak{h}}A)$ coincides so that $H^q(\frak{g}, \frak{h}; A)= 
H^q(\frak{g}/\frak{h},\, {\rm Inv}_{\frak{h}}A)$.
The definition of relative homology is similar to the definition of relative cohomology. For the space $C_q(\frak{g}, \frak{h}; A)$
of relative chain we must take $A\otimes_{\frak{h}}\Lambda^q(\frak{g}/\frak{h})$.

\subsection{The Hochschild-Serre spectral sequence}

\begin{theorem} \label{SH1}
\label{HStheorem}
{\rm (The Hochschild-Serre spectral sequence (see \cite{Fuks}, Section 1.5))}\, Let $A$ be a module over $\mathfrak g$. 
There exists a spectral sequence
$\{E_r^{p, q},\, d_r^{p, q}: \, E_r^{p, q} \rightarrow
E_r^{p+r, q-r+1}\}$
with the following properties: 
\begin{itemize}
\item{}
$E_1^{p, q} = H^q({\mathfrak h}, {\rm Hom}(\Lambda^p(\mathfrak{g}/\mathfrak{h}); {A}))$, \,\,\,\,\, $E_2^{p, 0}= H^p(\mathfrak{g},
\mathfrak{h}; {A})$.
\item{}
If $\mathfrak h$ is an ideal then $E_2^{p, q} = H^q({\mathfrak h}, H^q(\mathfrak{h},\, {A}))$.
\item{}
The term $E_{\infty}$ is associated to $H^*({\mathfrak g};\, {A}) =\bigoplus_qH^q({\mathfrak g};\, {A})$.
\item{}
The natural homomorphisms $H^q({\mathfrak g}; {A})\rightarrow H^q({\mathfrak h}; {A}),\, \, H^p({\mathfrak g}, {\mathfrak h}; 
{A})\rightarrow H^p({\mathfrak g}; {A})$ can be represented as compositions
$$
H^q({\mathfrak g}, {A}) \longrightarrow  E_{\infty}^{0, q}
\longrightarrow E_{1}^{0, q} = H^q({\mathfrak h};\, {A})
\,,\,\,\,\,\,
H^q({\mathfrak g}, \mathfrak{h}; {A})  =  E_{2}^{p, 0}
\longrightarrow E_{\infty}^{p, 0} \longrightarrow 
H^p({\mathfrak g}; {A})\,.
$$
\end{itemize}
If $A = \cA$ is an associative commutative algebra in which ${\mathfrak g}$ acts by means of derivations then the spectral sequence is multiplicative.
\end{theorem}
Some clarifications of this statement have to be made.
For the proof of the Theorem \ref{SH1} it is convenient to let
\begin{eqnarray}
F^pC^{p+q}(\mg; {A}) & = & \{c\in C^{p+q}(\mg; {A})\, \vert\, c(g_1, \ldots , g_{p+q}) = 0\,\,\,\,\,{\rm for}\,\,\,\,  g_1, \ldots , g_{p+q}
\in \mh\},
\label{Filt1}
\\
C^r(\mg; {A}) & = & F^0C^r(\mg; {A})\supset \ldots \supset F^rC^r(\mg; {A}) \supset F^{r+1}C^r(\mg; {A}) =0,
\label{Filt2}
\end{eqnarray}
where $dF^pC^{p+q}(\mg; {A})\subset F^pC^{p+q+s}(\mg; {A})$. Therefore, 
$\{F^p\}$ is a filtration in the complex $C^\bullet(\mg; {A})$.
Since there is a map $F^pC^{p+q}(\mg; {A})\rightarrow 
{\rm Hom}(\Lambda^p(\mg/\mh), {A})$ (which in fact is an epimorphism with kernel $F^{p+1}C^{p+q}(\mg; {A})$), we get the following isomorphism \begin{equation}
E_0^{p, q} = F^pC^{p+q}(\mg; {A})/F^{p+1}C^{p+q}(\mg; {A}) \longrightarrow 
C^q(\mh, {\rm Hom}(\Lambda^p(\mg/\mh)); {A}).
\end{equation}
This isomorphism commutes with differentials (see for detail \cite{Fuks}). Then we get 
\begin{eqnarray}
\!\!\!\!\!\!\!\!\!\!\!\!
E_1^{p, q} & = & H^q(\mh, {\rm Hom}(\Lambda(\mg/\mh)); {A}),
\\
\!\!\!\!\!\!\!\!\!\!\!\!
d_1^{p, q} :\,E_1^{p, q} & = & 
H^q(\mh, {\rm Hom}(\Lambda^p(\mg/\mh)); {A})
\longrightarrow E_1^{p+1, q} = H^q(\mh, {\rm Hom}
(\Lambda^{p+1}(\mg/\mh)); {A}). 
\label{dif} 
\end{eqnarray}
$d_1^{p,  q}$ is not induced by any natural homomorphism 
$\Hom (\Lambda^p({\mathfrak g}/{\mathfrak h}, {A}))\rightarrow 
\Hom (\Lambda^{p+1}({\mathfrak g}/{\mathfrak h}, {A}))$. But we do have the differential 
$
d: \, C^p({\mathfrak g}, {\mathfrak h}; {A})
\rightarrow C^{p+1}({\mathfrak g}, {\mathfrak h}; {A}),
$ 
which is a natural homomorphism
$
\Hom_{\mathfrak h}(\Lambda^p({\mathfrak g}/{\mathfrak h}), {A})
\rightarrow \Hom_{\mathfrak h}(\Lambda^{p+1}({\mathfrak g}/{\mathfrak h}), {A}).
$
The following diagram
\begin{equation}
\xymatrix@=10mm{
H^q({\frak h})\otimes C^p({\frak g}, {\frak h}; A) 
\ar[r]^{{\rm Id}\otimes d}
\ar@{=}[d]&H^q({\frak h})\otimes C^{p+1}({\frak g}, 
{\frak h}; A)
\ar@{=}[d]
\\
H^q({\frak h}, \Hom_{\frak h}(\Lambda^p({\frak g}/{\frak h}); A))
\ar@{->}[d]^{\rm inclusion}
&H^q({\frak h}, \Hom_{\frak h}(\Lambda^{p+1}({\frak g}/{\frak h}); A))
\ar@{->}[d]^{\rm inclusion}
\\
H^q({\frak h}, \Hom(\Lambda^p({\frak g}/{\frak h}); A))
\ar[r]^{d_1^{p, q}}
&H^q({\frak h}, \Hom(\Lambda^{p+1}({\frak g}/{\frak h}); A))
}
\label{diag1}  
\end{equation}
whose vertical arrows are induced by the inclusions $\Hom_{\frak h}(\Lambda^p({\frak g}/{\frak h}), A)\rightarrow
\Hom_{\frak h}(\Lambda^p({\frak g}/{\frak h}), A)$, is commutative.
In certain important cases vertical inclusions in (\ref{diag1})
turn out to be isomorphisms \cite{Fuks}; thus diargam (\ref{diag1}) is a satisfactory description of the differential $d_1^{p, q}$ as well as the term $E_2^{p, q}$.
In general  case these arrows are isomorphisms for $q=0$ \cite{Fuks}, so that
$
E_1^{p, q} = C^p({\frak g}, {\frak h}; A),\,
E_2^{p, 0} = H^p({\frak g}, {\frak h}; A).
$
The Theorem \ref{HStheorem} may be generalized to relative case. Indeed, if ${\frak k}$ is a subalgebra of the algebra 
${\frak h}$, there is a spectral sequence for which

- $E_1^{p, q} = H^q({\mh}, {\mathfrak k}, 
\Hom(\Lambda^p({\mg}/{\mh}); A))$\,,\,\,\,\,\,
$E_2^{p, 0} = H^p({\mg}, {\mh}; A)$\,.

- If ${\mh}$ is an ideal, then $E_2^{p, q} = H^p({\mg}/{\mh}, H^q({\mh}, {\mathfrak k}; A))$.

- The term $E_{\infty}$ is associated to 
$H^*({\mg}, {\mathfrak k}; A)$\,.


{\bf The infinite-dimensional $\mg$-module.} We can extend the Hochschild-Serre spectral sequence as follows 
(as proved by F. Williams in \cite{Williams80}). Let $\mg$ be a finite-dimensional Lie algebra over a base field $F$ of characteristic zero. Let $\mh \subset \mg$ be asubalgebra which is {\it reductive} in $\mg$ -- i.e. the adjoint action of $\mg$ on $\mh$ is semisimple.
Let $A$ be a $\mg$-module (i.e. we have a representation $\pi : \, \mg\rightarrow \mg\ml(A)$ of $\mg$ on $A$; one writes, as usual, $x\cdot v$ for the module structure, meaning $\pi(x) v$, for $(x, v) \in \mg\times A$). A might be infinite-dimensional but we assume that $A = \sum_r\oplus A_r$ is a direct sum, where each $A_r \subset A$ is finite-dimensional, $\mh$-invariant, and $\mh$-semisimple. 
\begin{theorem} The first- and second-order terms of the Hochschild-Serre spectral sequence generated by $\mh$ are given by
\begin{eqnarray}
E_1^{p, q} & \simeq &
H^q(\mh)\otimes \Lambda^p(\mg, \mh; A)\,,
\\
E_2^{p, q} & \simeq &
H^q(\mh)\otimes H^p(\mg, \mh; A)\,, 
\end{eqnarray}
exactly as in the case of a finite-dimensional $\mh$-semisimple $\mg$-module $A$ {\rm \cite{Hochschild53}}. Here, as usual, $H^q(\mh) = H^q(\mh, F)$ for $F$
the trivial $\mh$-module, and $H^p(\mg, \mh; A)$ is the relative Lie algebra cohomology, for the relative cochains $\Lambda^p(\mg, \mh; A)$
given by $\Lambda^p(\mg, \mh; A) = {\rm Hom}_{\mh}(\Lambda^p, \mg/\mh, A)$.
\end{theorem}
{\bf The Weyl DG-algebra.}
\label{Weyl}
Let us discuss a construction widely used in differential geometry, which yields a multiplicative complex possesing a series of 
supplementary structures and which is called the Weyl algebra \cite{Weyl}. 
Let ${\mathfrak h}$ be an arbitrary Lie algebra. Then the free DG-algebra\!
\footnote{Here and in the following DG-algebra means differential graded algebra.}
$W^{\bullet}({\mathfrak h})$
can be defined as a tensor product of symmetric and external algebras on the dual space of $\mathfrak h$. 
Suppose the homological powers of generators of external and symmetric algebras are equal to 1 and 2 respectively. 
We will use the standard notation from homological algebra for shifts of gradings in complexes $C[i]^j=C^{i+j}$. Then
\begin{equation}
W^{\bullet}({\mathfrak h}) = \Lambda^{\bullet}{\mathfrak h}^*\otimes
S^{\bullet}{\mathfrak h}^* = \Lambda^{\bullet}({\mathfrak h}^*[-1]
\otimes{\mathfrak h}^*[-2])\,,\,\,\,\,\,\,\,\,\,\,
W^{q}({\mathfrak h})=\bigoplus_{i+2j=q}
\Lambda^{i}{\mathfrak h}^*\otimes
S^{j}{\mathfrak h}^*\,.
\end{equation}
Choose the differential $d = d_1 + d_2$ on the Weyl algebra, where the differentials satisfy the Leibnitz rule. It is sufficient to define the action of differentials on generators.
Define the restrictions of $d_1$ and $d_2$ (Koszul differential) on generators:
$
d_1\mid_{\Lambda^1{\mathfrak h}^*}\,:\,\, \Lambda^1{\mathfrak h}^*
\rightarrow \Lambda^2{\mathfrak h}^*,\,\,
d_1\mid_{S^1{\mathfrak h}^*}\,:\,\, S^1{\mathfrak h}^*
\rightarrow \Lambda^1{\mathfrak h}^*\otimes 
S^1{\mathfrak h}^*;\,\,
d_2\mid_{\Lambda^1{\mathfrak h}^*}\,:\,\, \Lambda^1{\mathfrak h}^*
\stackrel{\rm Id}{\rightarrow} 
S^1{\mathfrak h}^*,\,\,
d_2\mid_{S^1{\mathfrak h}^*} = 0.
$
On the Weyl DG-algebra define a standard decreasing filtration
with respect to the differential $d_1$,
\begin{equation}
F^kW({\mathfrak h})\stackrel{{\rm def}}{=}\bigoplus_{2j\geq k}
\Lambda^{\bullet}{\mathfrak h}^*\otimes S^j{\mathfrak h}^*\,.
\end{equation}
Then the associated multiplicative spectral sequence 
$(EW_r^{p,q}({\mathfrak h}), d_r^{p,q})$ with $d_0= d_1$
has the following initial terms:
\begin{eqnarray}
EW_0^{p,q}({\mathfrak h}) =  \left\{\begin{array}{ll}
\!\Lambda^q{\mathfrak h}^*\otimes 
S^{\frac{p}{2}}{\mathfrak h}^*\,
\, &{\rm if}\,\,\,\,\,p\,\,\,{\rm even}
\\
\!0 
&{\rm if}\,\,\,\,\,p\,\,\,{\rm odd}
\end{array}\right.,
\,\,\,\,\,
EW_{r=1, 2}^{p,q}({\mathfrak h})
=  
\left\{\begin{array}{ll}
\!H^q({\mathfrak h}, S^{\frac{p}{2}}{\mathfrak h}^*)\,\,\,
&{\rm if}\,\,\,\,\,p\,\,\,
{\rm even}
\\
\!0\,\,
&{\rm if}\,\,\,\,\,p\,\,\,{\rm odd}
\end{array}\right.
\label{SS1}
\end{eqnarray}
The spectral sequence (derived from the filtration of the external algebra) degenerates at the first term, 
which prooves that the Weyl DG-algebra is acyclic.

{\bf The relative Weyl algebra.}
Let $\mathfrak g$ be a Lie algebra, 
$\mathfrak h\subset \mathfrak g$ a finite dimensional subalgebra. 
The relative Weyl algebra is the commutative DG-algebra
\begin{equation}
W(\mathfrak g,\mathfrak h)=\bigoplus_{i,j\geq0} W^{i,j}(\mathfrak g,\mathfrak h),\,\,\,\,\,
{\rm where}\,\,\,\,\, W^{i,j}(\mathfrak g,\mathfrak h)
=[\Lambda^i(\mathfrak g/\mathfrak h)
\otimes S^j\mathfrak g]^{*\mathfrak h}\,\,\,\,\,
{\rm has}\,\,\, {\rm grading}\,\,\, i+2j\,.
\end{equation} 
The relative Weyl algebra $W^{\bullet}({\mathfrak g}, {\mathfrak h})$ is the Weyl DG-algebra $W^{\bullet}({\mathfrak g})$ 
such that an element $c$ in $\Lambda^q{\mathfrak g}^*\otimes S^p{\mathfrak g}^*
= {\rm Hom}(\Lambda^q{\mathfrak g}, S^p{\mathfrak g}^*)\subset W^{\bullet}({\mathfrak g})$ is an element of relative Weyl algebra
which satisfies the following conditions:
\begin{equation}
c(g_1,g_2,...,g_q)=0 \,\,\,\, {\rm if}\,\,\,\, g_1\in {\mathfrak h}\,;\,\,\,\,\,\,\,\,\,\,
(\delta_1c)(g_1,g_2,...,g_q, g_{q+1})=0 \,\,\,\, {\rm if} \,\,\,\,g_1\in {\mathfrak h}\,.
\end{equation}
If ${\mathfrak h}= {\mathfrak g}$, then $W^{2p}({\mathfrak g}, {\mathfrak g})=[S^p{\mathfrak g}^*]^{\mathfrak g}$,\,
$W^{2p+1}({\mathfrak g},{\mathfrak g})=0$ and this complex coincides with its cohomology.
The standard filtration on all Weyl algebra transfers to a relative Weyl subalgebra. In addition the spectral sequence has 
the following initial terms:
\begin{equation}
EW_{r=1, 2}^{p,q}({\mathfrak g}, {\mathfrak h})= 
\left\{\begin{array}{ll}
H^q({\mathfrak g}, {\mathfrak h}; S^{\frac{p}{2}}{\mathfrak g}^*)\,
& {\rm if}\,\,\,\,\,p\,\,\,{\rm even}
\\
0\,
& {\rm if}\,\,\,\,\,p\,\,\,{\rm odd}
\end{array}\right.
\label{SS2}
\end{equation}
The spectral sequence (derived from the filtration of the exterior algebra) degenerates at the first term, and its total cohomologies are:
$
H^p(W({\mathfrak g}, {\mathfrak h})) = [S^{\frac{p}{2}}{\mathfrak h}^*]^{\mathfrak h}
$
\,if \,$p$\,\, even, and 
$
H^p(W({\mathfrak g}, {\mathfrak h})) = 0
$
\,if\, $p$ odd.

\section{Appendix B: Sheaves and categories}

{\bf A preheaf.} A {\it presheaf} ${\cF}$ on topological space $X$ consists of the following data:

For every open set $U\subset X$ one can associate an abelian
group ${\cF}(U)$.

If $A\subset U$ are open sets then one has a restriction
homomorphism $r_{UA}: {\cF}(U)\rightarrow {\cF}(A)$.

\noindent
The following conditions hold: ${\cF}(\emptyset) = 0$;
\, $r_{UU}$ is the identity map;\,
if $W\subset A \subset U$
then $r_{AW}r_{UA}=r_{UW}$. 
For the restriction $r_{UA}({\cR})$,
${\cR} \in {\cF}(U)$ we use the notation ${\cR}|_{A}$.
An element of ${\cF}(U)$ is called a section of ${\cF}$
over $U$, while an element of ${\cF}(X)$ is called a global
section.

\noindent
{\bf A sheaf.}
A presheaf ${\cF}$ is called {\it sheaf} if for every
collection $U_{j}$ of open subsets of $X$ with
$U=\bigcup \limits_{j\in I}U_{j}$ the following
axioms hold:

- If ${\cR}, {\cQ} \in {\cF}(U)$ and
$r_{UU_{j}}({\cR})=r_{UU_{j}}({\cQ})$ $\forall$
$j$, then ${\cR}={\cQ}$.

- If ${\cR}_{j}\in {\cF}(U_{j})$
and if for $U_{j}\bigcap U_{j}\neq \emptyset$,
$r_{U_{j},U_{j}\bigcap U_{\ell}}({\cR}_{j})=
r_{U_{\ell},U_{j}\bigcap U_{\ell}}({\cR}_{\ell})$
$\forall j$, then there exists an
${\cR}\in {\cF}(U)$ such that $r_{U,U_{j}}({\cR})
={\cR}_{j}$, $\forall j$.

Let ${\cF}$ and ${\cE}$ be presheaves
over $X$. Then a morphism of presheaves
$\alpha: {\cF}\rightarrow {\cE}$ is a collection of maps
$\alpha(U): {\cF}(U)\rightarrow {\cE}(U)$,
satisfying the relation $r_{UA}\alpha(U)=\alpha(A)r_{UA}$.
Morphisms of sheaves are morphisms of the underlying presheaves.

\noindent
{\bf Coherent sheaves.}
Let $X$ be a complex manifold. A sheaf
${\cF}$ over $X$ is called a {\it coherent sheaf} of
${\cO}$-modules if for each $z \in X$ there is a neighborhood
$U$ of $z$ such that there is an exact sequence of sheaves over $U$:
$$
0\longrightarrow {\cF}|_{U}\longrightarrow
{\cO}^{\oplus p_{1}}|_{U}
\longrightarrow {\cO}^{\oplus p_{2}}|_{U}
\longrightarrow \ldots \longrightarrow
{\cO}^{\oplus p_{j}}|_{U}
\longrightarrow 0\,
$$

\noindent
{\bf Categories.}
A category $\cC$ consists of the following data:
a class Ob $\cC$ of objects $A$, $B$, $C$, $\ldots$;
a family of disjoint sets of morphisms 
Hom($A$, $B$), one for each ordered pair $A$, $B$ of objects;
a family of maps
$
\mbox{Hom}(A,B)\times \mbox{Hom}(B, C)\rightarrow
\mbox{Hom}(A, C),
$
one for each ordered triplet $A$, $B$, $C$ of objects.
These data obey the axioms:

- If $f : A\rightarrow B,
g : B\rightarrow C,\, h : C\rightarrow D,$
then composition of morphisms is associative,
that is, $h(gf)= (hg)f$.

- To each object $B$ there
exists a  morphism ${\id}_{B}: B\rightarrow B$ such that 
${\id}_{B} f=f$\ , $g {\id}_{B}=g$ for $f:
A\rightarrow B$ and $g: B\rightarrow C$.

{\bf Additive category.} An additive category is a category in which each set of morphisms 
Hom$(A, B)$ has the structure of an abelian group. The following axioms hold:

- Composition of morphisms
is distributive:
$
(g_{1}+g_{2})f = g_{1}f
+g_{2}f\ , \ \ \ h(g_{1}
+g_{2})=hg_{1}+ hg_{2}
$
for any $g_{1}, g_{2} : B \rightarrow C\ , \ \ f : A\rightarrow
B\ ,\ \ h : C\rightarrow D$.

- There is a null object $0$ such that Hom($A$; $0$)
and Hom($0$;~$A$) consist of one
morphism for any $A$.

- To each pair of
objects $A_{1}$ and $A_{2}$ there
exists an object $B$ and four morphisms
$
A_1\stackrel{j_1}\rightarrow B\stackrel{\ell_2}\rightarrow
A_2\stackrel{j_2}\rightarrow B\stackrel{\ell_1}\rightarrow A_1
$,
which satisfy the identities
$
\ell_kj_k = {\id}_{A_k}, \, (k=1,2), \,\, j_1\ell_1+j_2\ell_2 = 
{\id}_B, \,\, \ell_2j_1=\ell_1j_2 =0.
$

{\bf Abelian category.}
It is an additive category $\cC$
which satisfies the additional axiom:

- To each morphism $f : A\rightarrow B$ there
exists the sequence
$
K\ \stackrel{k}{\rightarrow}\
A \ \stackrel{i}{\rightarrow}\
I\ \stackrel{j}{\rightarrow}\ B \
\stackrel{c}{\rightarrow}\ K^{'}
$
with the properties: $ji=f$; $K$ is a kernel
of $f$, \, $K^{'}$ is a cokernel of $f$;
$I$ is a cokernel of $k$ and a kernel of $c$.

The category of coherent sheaves is an abelian category $\cC$.

\noindent
{\bf Derived category.}
The definition of the derived category
D(${\cC}$) proceeds as follows \cite{Gelfand}:
\begin{enumerate}
\item{} We begin with the category of complexes of coherent
sheaves {\bf Kom}(${\cC}$): Ob {\bf Kom}(${\cC}$) =
$\{$complexes $\cE^{\bullet}$ of coherent sheaves $\}$;
Hom($\cE^{\bullet}$, $\cF^{\bullet}$)
= morphisms of complexes $\cE^{\bullet} \rightarrow
\cF^{\bullet}$. 
\item{} The homotopy category {\bf K}(${\cC}$)
can be determined as follows: Ob {\bf K}(${\cC}$) = 
Ob {\bf Kom}(${\cC}$), Mor {\bf K}(${\cC}$) = 
Mor {\bf Kom}(${\cC}$) modulo homotopy equivalence.
\item{} Finally the derived category D$({\cC})$ 
is determined as follows:
Ob D$({\cC})$ = Ob {\bf K}(${\cC}$).
\end{enumerate}
The morphisms of D(${\cC}$)
are obtained from morphisms in ${\bf K}({\cC})$ by inverting
all quasi-isomorphisms. The derived category
D$({\cC})$ is an additive category.

\section{Appendix C: The HKR isomorphism}

Let $\Delta: X\rightarrow X^2$ be the diagonal embedding.
There exists a quasi-isomorphism \cite{Hochschild,Kontsevich,Swan,Yekutieli}:
\begin{equation} 
I: \,\Delta^*\cO_\Delta \stackrel{\sim}{\longrightarrow}
\bigoplus_i \Omega_X^i[i]\,.
\end{equation}
Here the right hand side denotes the complex whose $i$-th term is $\Omega_X^i$, and all 
differentials are zero.

A sketch of a proof of this statement is as follows. 
Recall that if $\cA$ is a commutative $K$-algebra 
there exists a standard resolution of $\cA$ as an $\cA^e = \cA\otimes_
{K} \cA$-module. These are $\cA$ bimodules with the bimodule structure given by the 
multiplication by $\cA$ on the left and right copies of $\cA$. It is clear that these are 
projective bimodules. We can denote an element of $\cA^{\otimes n}$ as 
$[a_1|a_2| \ldots |a_n]$. Let ${\mathfrak B}_i(\cA) = \cA^{\otimes (i+2)}$, $i\geq 0$, 
where the tensor product is taken over $K$ (it is an $\cA^e$-module by multiplication in the 
first and last factor).  Then the bar resolution is defined to be the complex of $\cA^e$-modules
\begin{equation}
\cdots\longrightarrow {\mathfrak  B}_i(\cA) 
\stackrel{\partial}{\longrightarrow} \cdots 
\stackrel{\partial}{\longrightarrow} {\mathfrak B}_2(\cA) 
\stackrel{\partial}{\longrightarrow}{\mathfrak B}_1(\cA) 
\stackrel{\partial}{\longrightarrow}{\mathfrak B}_0(\cA)
\longrightarrow \cA \longrightarrow 0\,.
\label{complex1}
\end{equation}
Here the differential $\partial$ is the linear $\cA^e$-morphism given by the formula 
\begin{equation} 
\partial(a_0\otimes \cdots\otimes a_{q+1}) =  
\sum_{i=0}^q(-1)^ia_0\otimes \cdots \otimes a_ia_{i+1}
\otimes\cdots\otimes a_{q+1}\,.
\end{equation}
The complex (\ref{complex1}) is split-exact with splitting homomorphism 
$\sigma (a_0\otimes \cdots a_{q+1}) =  
a_0\otimes \cdots a_{q+1}\otimes \id$. The homomorphism $\sigma$
is $\cA$-linear when $\cA$ acts via $a\mapsto a\otimes \id$ \cite{Loday}. 

If $X$ were affine then $X={\rm Spec}\,\cA $ and we could use the above resolution to 
compute $\Delta^* \cO_\Delta$. Namely, $\cA$
can be viewed as an $\cA^e = \cO_{X^2}$-module, and the modules ${\mathfrak B}_i$ are 
$\cA^e$-flat. The complex obtained by tensoring the bar resolution over $\cA^e$ with $\cA$ 
is called the bar complex,
\begin{equation}
\cdots\longrightarrow {C}_i(\cA) \longrightarrow \cdots \longrightarrow {C}_1(\cA) 
\longrightarrow {C}_0(\cA) 
\longrightarrow 0\,,\,\,\,\,\,\,\,\,\,
{C}_i(\cA) = {\mathfrak B}_i(\cA) \otimes_{\cA^e} \cA\,,
\end{equation}
where ${C}_i(\cA)$ is the module of any degree $i$ Hochschild chains of $\cA$, and the 
differential is obtained from the differential $\partial$ of (\ref{complex1}).
However when one tries to sheafify the bar resolution to
obtain a complex of sheaves on a scheme, the resulting sheaves are ill-behaved 
(not quasi-coherent).
As a replacement, one can use the complete bar resolution
\cite{Yekutieli}.  

For $i\geq 0$, let ${\mathfrak X}^i$ be the formal completion of the scheme 
${\mathfrak X}^i = X\times
\cdots\times X$ along the diagonal embedding of $X$. Define
for any $i\geq 0$, $\widehat{{\mathfrak B}}_i(X) := 
\cO_{{\mathfrak X}^{i+2}}$ (which is a sheaf of abelian groups on the topological space $X$). 
The sheaf of degree $i$ complete Hochschild chains of $X$ is 
$\widehat{C}_i:= \widehat{{\mathfrak B}}_i(X)\otimes_{\cO_{{\cX}^2}}\cO_X$.
One can formally complete and sheafify the original bar resolution to get the complete bar 
resolution 
\begin{equation}
\cdots\longrightarrow \widehat{{\mathfrak B}}_i(X) 
\longrightarrow \cdots 
\longrightarrow \widehat{{\mathfrak B}}_1(X) 
\longrightarrow \widehat{{\mathfrak B}}_0(X) 
\longrightarrow 0\,,
\end{equation}
where the maps are locally obtained from the maps of the original bar complex. As a result, 
the complete bar resolution is an exact resolution of $\cO_\Delta$ by sheaves of flat 
$\cO_{X^2}$-modules. So over an affine open set 
$U = {\rm Spec}\,\cA$ of $X$, 
$\Gamma(U, \widehat{{\mathfrak B}}_i(X))$ is the completion $\widehat{{\mathfrak B}}_i(\cA)$ 
of ${\mathfrak B}_i(\cA)$ at the ideal $I_i$ given by the kernel of the
multiplication map ${\mathfrak B}_i(\cA) = \cA^{\otimes i} \rightarrow \cA$.

Let us consider the complete bar resolution as a flat resolution of $\cO_\Delta$ on $X^2$, 
and let us compute
$\Delta^*\cO_\Delta$. This means that we are looking at the tensoring of the complete bar 
resolution over $\cO_{X^2}$ with $\cO_\Delta$. 
Then the resulting complex is called the complex of complete Hochschild chains of $X$,
\begin{equation}
\cdots\longrightarrow \widehat{C}_i(X) 
\longrightarrow 
\cdots \longrightarrow \widehat{C}_1(X) 
\longrightarrow \widehat{C}_0(X) 
\longrightarrow 0\,,\,\,\,\,\,\,\,\,\,\,\,\,
\widehat{C}_i(X) = \widehat{{\mathfrak B}}_i(X) \otimes_{\cO_{X^2}} \cO_\Delta\,.
\end{equation}
Over any affine open $U= {\rm Spec}\, \cA$ define the maps
$I_i:  {C}_i(\cA) \rightarrow \Omega_{\cA/k}^i$ by setting 
\begin{equation}
I_i(({\id}\otimes a_1\otimes\cdots\otimes a_i 
\otimes {\id}) \otimes_{\cA^e}{\id}) = d a_1\wedge da_2 
\wedge\cdots \wedge da_i\,.
\end{equation}
These maps are continuous with respect to the topology \cite{Yekutieli} and they can be 
completed and sheafified to maps
$I_i:\widehat{C}_i(X) \rightarrow \Omega_X^i.
$
These maps also commute with the zero differentials of the complex $\oplus_i \Omega_X^i$, 
so they extend to a morphism of complexes
$I:\Delta^*\cO_\Delta \rightarrow \bigoplus_i \Omega_X^i[i]
$
which can be seen to be a quasi-isomorphism in characteristic
0 \cite{Yekutieli}. In the affine case this is essentially the HKR theorem~\cite{Hochschild}.

\noindent
{\bf The Hochschild structure.} In the case of a quasi-projective variety $X$ the HKR 
isomorphism induces isomorphisms of graded vector spaces
$
I^{\rm HKR} : \, HH^k(X) \stackrel{\sim}{\rightarrow} HT^k(X),\,
$
$
I_{\rm HKR}  : \, HH_k(X) \stackrel{\sim}{\rightarrow} H\Omega_k(X)
$:
\begin{eqnarray}
HH^k(X) & = & \Hom_{X^2}(\cO_\Delta, \cO_\Delta[k]) \cong
\Hom_X(\Delta^*\cO_\Delta, \cO_X[k]) 
\cong  \Hom_X(\bigoplus_i \Omega_X^i[i], \cO_X[k]) 
\nonumber \\
& = & \bigoplus_i H^{k-i}(X, \Lambda^i TX) = HT^k(X)\,,
\label{HH0}
\\
HH_k(X) & = & \Hom_{X^2}(\Delta_!\cO_X[k],\cO_\Delta) \cong
\Hom_X(\cO_X[k], \Delta^*\cO_\Delta) 
\cong  \Hom_X(\cO_X[k], \bigoplus_i \Omega_X^i[i]) 
\nonumber \\
& = & \bigoplus_i H^{i-k}(X, \Omega^i_X) = H\Omega_k(X)\,.
\end{eqnarray}
As a result, the Hochshild structure consists of \cite{Caldararu1,Caldararu2}:

- A graded ring $HH^i(X)$, the Hochschild cohomology ring, defined as
$
HH^i(X) = \\
\Hom_{{\rm D}^b_{\rm coh}(X^2)}(\cO_\Delta, \cO_\Delta[i]),
$

- A graded left $HH^i(X)$-module $HH_i(X)$, the Hochschild homology module, defined as \\
$
HH_i(X) = \Hom_{{\rm D}^b_{\rm coh}(X^2)} (\Delta_!\cO_X[i], \cO_\Delta)\,,
$
and

- A non-degenerate pairing $\langle\cdot,\cdot\rangle$
defined on $HH_i(X)$, the generalized Mukai pairing,

- The connection between Hochshild and harmonic structures is given by the HKR isomorphism.

\end{document}